\documentclass[sn-basic]{sn-jnl}

\makeatletter
\let\old@oddsidemargin\oddsidemargin
\let\old@evensidemargin\evensidemargin
\makeatother

\usepackage{graphicx}%
\usepackage{multirow}%
\usepackage{amsmath,amssymb,amsfonts}%
\usepackage{amsthm}%
\usepackage{bm}%
\usepackage{mathrsfs}%
\usepackage[title]{appendix}%
\usepackage{xcolor}%
\usepackage{tikz}
\usetikzlibrary{positioning,shapes,arrows,calc,fit,backgrounds}
\usepackage{pgfplots}
\pgfplotsset{compat=1.18}
\usepackage{textcomp}%
\usepackage{manyfoot}%
\usepackage{booktabs}%
\usepackage{algorithm}%
\usepackage{algorithmicx}%
\usepackage{algpseudocode}%
\usepackage{listings}%
\usepackage{tikz}%
\usetikzlibrary{positioning,shapes,arrows}%
\usepackage{subcaption}%
\usepackage{lineno}

\newcommand{\vx}{\mathbf{x}}
\newcommand{\vy}{\mathbf{y}}
\newcommand{\valpha}{\boldsymbol{\alpha}}

\definecolor{darkblue}{RGB}{0,51,102}
\definecolor{lightblue}{RGB}{230,240,250}
\definecolor{accent}{RGB}{204,0,0}
\definecolor{greenaccent}{RGB}{0,153,51}

\theoremstyle{thmstyleone}%

\theoremstyle{thmstyletwo}%
\theoremstyle{thmstylethree}%
\unnumbered 

\begin{document}

\title[Cohort-amortized personalization]{
Cohort-amortized personalization: navigating the privacy-utility frontier for
virtual brain twins.
}


\author[1]{\fnm{Amirhossein} \sur{Esmaeili$^\ddagger$}}\email{amirhossein.esmaeili@univ-amu.fr}
\author[1]{\fnm{M Marmaduke} \sur{Woodman$^{*,\ddagger}$}}\email{marmaduke.woodman@univ-amu.fr}
\author[1]{\fnm{Nina} \sur{Baldy}}\email{nina.baldy@univ-amu.fr}
\author[1]{\fnm{Abolfazl} \sur{Ziaeemehr}}\email{abolfazl.ziaee-mehr@univ-amu.fr}

\author[1,2]{\fnm{Julia} \sur{Makhalova}}\email{julia.scholly@ap-hm.fr}
\author[1]{\fnm{Huifang} \sur{Wang}}\email{huyfang.wang@univ-amu.fr}

\author[3]{\fnm{Daniele} \sur{Marinazzo} \email{daniele.marinazzo@ugent.be}}
\author[4,5]{\fnm{Svenja} \sur{Caspers} \email{svenja.caspers@hhu.de}}
\author[1,2]{\fnm{Fabrice} \sur{Bartolomei}}\email{fabrice.bartolomei@ap-hm.fr}

\author[1]{\fnm{Meysam} \sur{Hashemi$^\dagger$}}\email{meysam.hashemi@univ-amu.fr}
\author[1]{\fnm{Viktor} \sur{Jirsa$^{\dagger}$}}\email{viktor.jirsa@univ-amu.fr}

\affil[1]{\orgdiv{Institut de Neurosciences Des Syst\`emes},
\orgname{Aix Marseille Universit\'e, INSERM},
\orgaddress{Marseille, 13005, France}}

\affil[2]{
    \orgdiv{Epileptology and Clinical Neurophysiology Department, Timone Hospital},
    \orgname{APHM},
    \orgaddress{Marseille 13005, France}
}

\affil[3]{%
    \orgdiv{Department of Data Analysis},
    \orgname{University of Ghent},
    \orgaddress{B-9000 Ghent, Belgium}%
}

\affil[4]{
    \orgdiv{Institute for Anatomy I}, 
    \orgname{University Hospital Düsseldorf \& Heinrich Heine University} 
    \orgaddress{Düsseldorf, Düsseldorf, Germany}
}

\affil[5]{
    \orgdiv{Institute of Neuroscience and Medicine (INM-1)}, 
    \orgname{Research Centre Jülich},
    \orgaddress{Jülich, Germany}
}



\abstract{
  Personalized generative brain models require individual neuroimaging data
  that privacy constraints and re-identification risk make difficult to share,
  while per-subject fitting procedures cost hours of compute---limiting clinical
  translation and multi-site collaboration. We introduce cohort-amortized
  personalization (CAP), which replaces \textit{data sharing} with
  \textit{model sharing}: a neural density estimator is trained on
  simulations from a mechanistic whole-brain model under a low-rank cohort
  prior, and only the compact estimator is distributed, so new subjects are
  personalized in seconds on their own data alone. To make this prior both
  compact and atlas-independent, a cross-atlas autoencoder (CrossCoder)
  maps connectomes from 20 anatomical atlases into a shared latent space,
  enabling deployment across sites with heterogeneous atlases. We
  validate CAP on two cohorts: 21 patients with drug-resistant epilepsy
  (epileptogenic-zone localization F1\,=\,0.56) and 832 subjects from the
  1000BRAINS aging cohort (predicted age r\,=\,0.44); in both, CAP matches or
  exceeds per-subject inference with hours-to-seconds speed-up. Because the
  shared artifact couples a cohort prior to a mechanistic simulator, it can serve
  as a \textit{mechanistic surrogate} supporting in-silico experimentation and
  synthetic-cohort generation without raw-data access---a governance-audited
  alternative we term \textit{synthetic access},
  allowing for wider adoption of personalized modeling in more diverse settings.
}

\keywords{privacy-utility, virtual brain twin, amortized inference, personalized medicine, epilepsy, aging}



\maketitle

{\footnotesize
\noindent$^\ddagger$These authors contributed equally to this work as first authors.\\
\noindent$^\dagger$These authors contributed equally to this work as senior authors.\\
\noindent$^*$Corresponding author: Marmaduke Woodman. Email: marmaduke.woodman@univ-amu.fr
}

\section{Introduction}\label{intro}


    Neuroimaging cohort data require difficult compromises between privacy and utility. Structural and functional neuroimaging, after anonymization still permit re-identification of individuals within cohorts.  Under EU General Data Protection Regulation, these are personal data which present legal and technical challenges for research infrastructures \citep{Bajada2025, Voigt2017}. Structural and functional connectomes, moreover, carry individual-specific signatures that enable re-identification even from small subsets of
    connectivity patterns \citep{Smith2015, Finn2015, MirandaDominguez2014, Byrge2019, Wachinger2015}, and generative re-identification attacks suggest that pseudonymization alone may not eliminate privacy risk in incomplete
    datasets~\citep{Rocher2019}. In parallel, advances in neurotechnology \& AI
    enabling (a) decoding brain activity from recorded signals \citep{Yuste2023} and (b) neural profiling based on social media content \citep{dascoli2026foundationmodelvisionaudition, dascoli2024decodingindividualwordsnoninvasive} amplify privacy concerns, largely sidestepped by effectively unregulated technology platforms \citep{doctorow2025enshittification}. The intersection of governance frameworks, including the EU AI Act and the NIST AI Risk Management Framework, with neuroinformatics incentivizes the development of privacy-preserving techniques for governance at scale without hampering fundamental research \citep{Alhejri2024}.

    The consequence for neuroscience is that personalized generative models,
    which require access to individual neuroimaging data are difficult to develop or deploy on open research infrastructures without controlled-access arrangements. So, while whole-brain modeling, in which neural mass models of local dynamics are coupled through subject-specific structural connectomes, has emerged as a powerful framework for constructing personalized virtual brain twins to understand and predict brain dynamics in health and disease \citep{Jirsa2023, Patow2024, Wang2023,Wang2025, Hashemi2025}, the inference procedures required to fit these models, be it Markov chain Monte Carlo variants \citep{Carpenter2017} or approximate Bayesian methods \citep{Tejero-Cantero2020}, are only effective with significant computational resources, a prohibitive requirement for translation of personalized models built with neuroimaging cohorts to routine clinical practice. This key tension between privacy and utility limits the downstream impact of neuroimaging cohort modeling.

    We obviate this tension by integrating amortized inference \citep{Hashemi2023,Baldy2025} with mechanistic brain models: a low-rank probabilistic model of the cohort data is derived and used as the prior distribution, a mechanistic whole-brain model predicting features from parameters is simulated, and an amortized density estimator learns to map features to parameters. 
    
\begin{figure}[t]
        \centering
        \includegraphics[width=\textwidth]{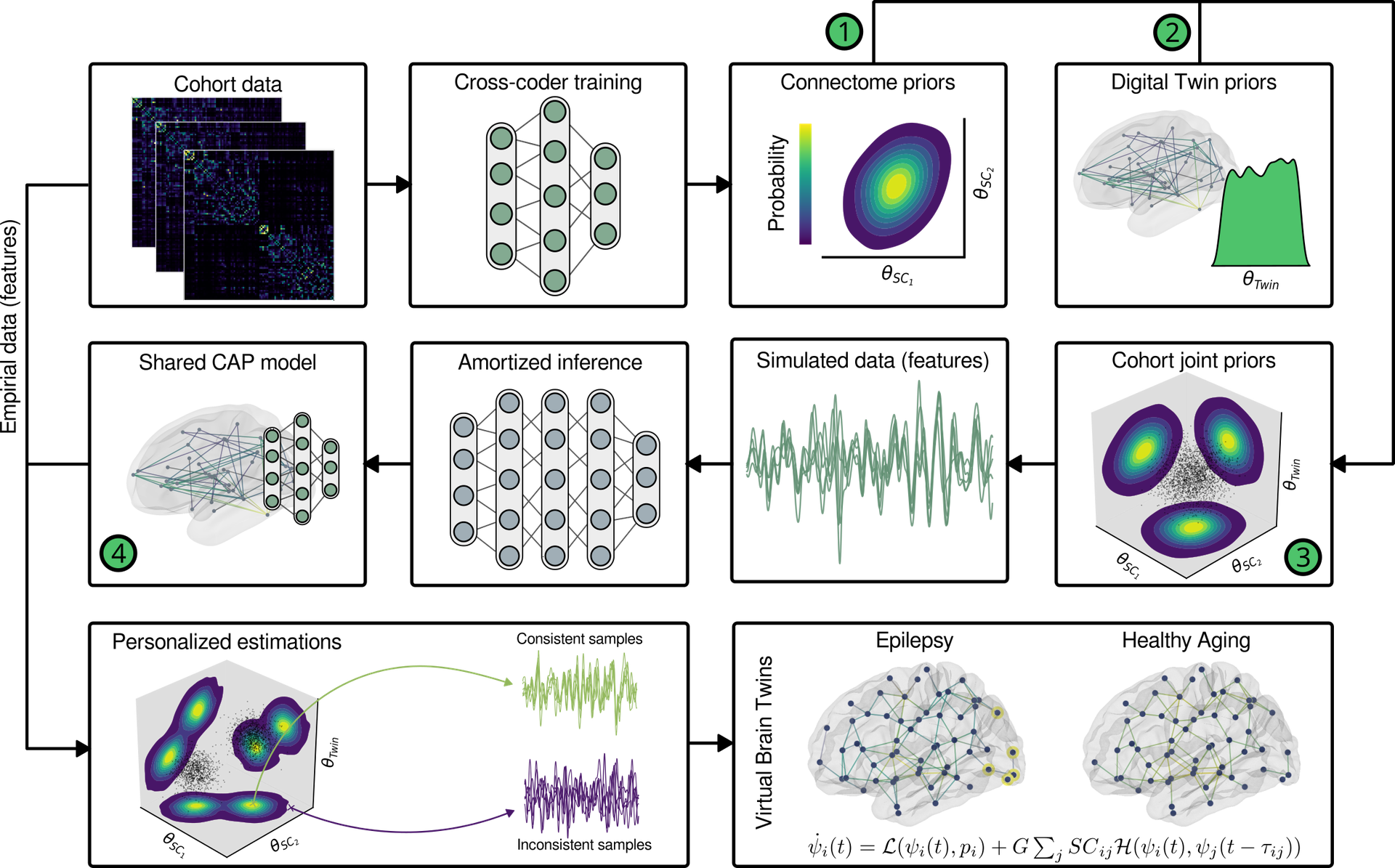}
        \caption{
          Basic cohort-amortized personalization (CAP) workflow. (1) Cohort data are used to derive a low-rank prior which contains no raw individual data and is intended to be shareable, while capturing generative statistics over the cohort. (2) Clinical hypotheses drive the configuration of empirically-informed virtual brain twin priors. (3) The priors in the previous subsystems are coalesced into a joint prior distribution to form a privacy-preserving data-twin representation of the cohort. (4) The shared CAP model serves as the core component of the workflow and is trained using cohort-level data. Personalized application for each subject is illustrated in blue in the bottom row. Empirical data features are used as inputs to the shared CAP model, which outputs subject-specific personalized parameters. We demonstrate the clinical applications of this framework in studies of epilepsy and healthy aging.
        }
        \label{fig:cap_workflow}
    \end{figure}

    The key insight for privacy is that the probabilistic model of the cohort data
    is anonymous, and during this offline training phase the compute-intensive simulation process can therefore take place in non-secure compute environments. Once training is complete, the resulting estimator is small and efficient to run, in addition to being anonymous.
    Personalization occurs only when the estimator is applied to their own data, so the shared estimator contains no raw individual data and is designed to be shareable without transferring patient records. We refer to this approach as cohort-amortized personalization (CAP), in which the shared model functions as a \textit{generative surrogate}: it may encode the statistical structure of the cohort \emph{and} the equations that map personal data to personal brain dynamics (Figure \ref{fig:cap_workflow}). Remote researchers may therefore generate synthetic cohorts, run virtual lesions or stimulation experiments, and benchmark new methods against a mechanistic ground truth without needing to navigate data transfer agreements or access agreements for secure compute environments. CAP enables \textit{synthetic access}, in which the personal data stay local, the pretrained model travels, and the science proceeds within the boundaries of what the generative family can faithfully represent \citep{SanzLeon2013, Dollomaja2025, Vaden2020}.

    This work demonstrates CAP first through a generic synthetic simulation scenario with Human Connectome Project (HCP) \citep{van2013wu} and 1000BRAINS \citep{Caspers2014} structural connectomes preprocessed in 20 atlases \citep{popovychHCP, popovych1000brains} for illustration purposes, followed by two cohort specific models. The first cohort, comprised of 21 patients with drug-resistant epilepsy, leads to a shareable cohort model where personalized regional variability predicts seizure initiation and propagation, validated by clinical experts  \citep{Wang2023,Wang2025,Jirsa2023}. The second cohort, comprised of a larger set of 832 subjects from the 1000BRAINS imaging study \citep{Caspers2014} spanning the adult lifespan, leads to a shareable cohort model where structural disconnection accounts for a modified resting state, with validation against previously per-subject modeling results \citep{Lavanga2023}. Both cohort models demonstrate personalization in seconds rather than hours, and the resulting models can be shared without exposing individual patient data. We conclude by discussing the results, related work and future extensions.

\section{Results}\label{sec:results}


\subsection{Cohort distributions allow amortized personalization}
  \label{sec:crosscoder_validation}

%
%
%
%

\begin{figure}[ht]
    \centering
    \includegraphics[width=\textwidth]{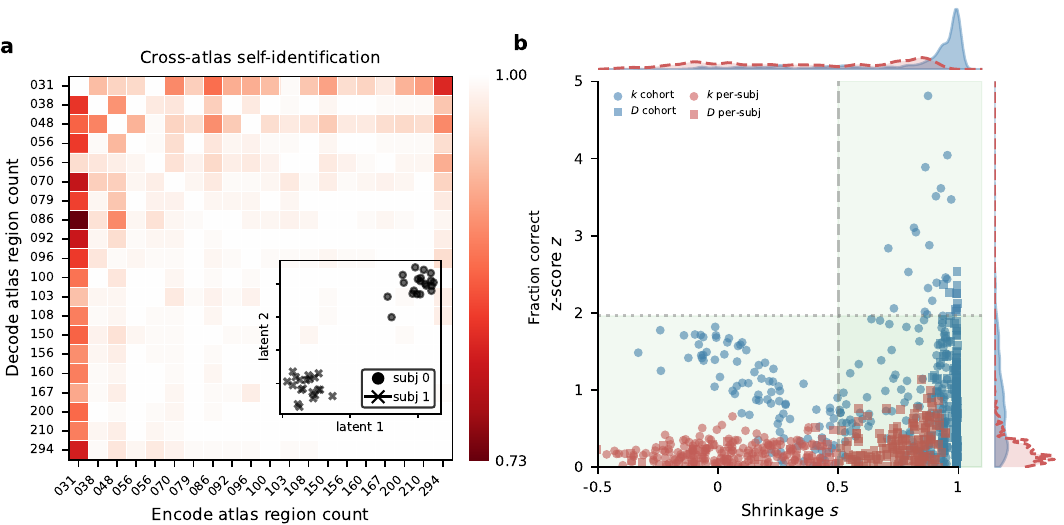}
    \caption{
        CrossCoder identifiability and synthetic-data SBI calibration.
        \textbf{(a)~Cross-atlas self-identification.}
        Each cell $(i,j)$ of the matrix gives the fraction of subjects whose
        latent vector for the \emph{encode} atlas $i$ is closest to the
        latent vector of the \emph{decode} atlas $j$ (or to itself on
        the diagonal). The colormap saturates to white at $1.0$, so the
        near-perfect diagonal self-identification and the dense band of
        high cross-atlas agreement read as whitespace, leaving only the
        small set of worse-performing atlas pairs (down to $0.73$) shaded.
        \emph{Inset:} latent-space weights for two representative subjects
        (o and x, 20 atlases), showing clear and
        consistent subject-level separation in the shared
        $n_{\mathrm{lat}}=16$ latent space.
        \textbf{(b)~Cohort vs.\ per-subject SBI calibration.}
        Joint scatter of posterior shrinkage $s$ (x) vs.\ z-score $z$ (y)
        for coupling $k$ (o) and noise amplitude $D$ (box); cohort posteriors
        are filled blue, per-subject posteriors red, overlaid with
        $\alpha=0.6$. Top and right marginals give the pooled-$k{+}D$ KDE
        per group (cohort: filled blue; per-subject: dashed red;
        per-subject shrinkage outliers $s < -0.5$ are clipped to the view).
        High shrinkage ($s > 0.5$) indicates concentrated posteriors and
        low z-score ($z < 1.96$) indicates well-calibrated uncertainty.
        Noise $D$ is near-universally well identified
        ($s_D \approx 0.98$); coupling $k$ has moderate but variable
        identifiability ($s_k \approx 0.57 \pm 0.35$).
    }
    \label{fig:vcc-combo}
    \label{fig:vcc-valid}
    \label{fig:vcc-synth}
    \label{fig:vcc_confusion}
    \label{fig:vcc_latent_weights}
    \label{fig:shrinkage-zscore}
    \label{fig:cohort-calibration}
\end{figure}

  \textbf{CrossCoder and cohort-amortized inference.}
    Structural connectomes serve as a key personalization for virtual brain twin models, yet amortized simulation based inference (SBI) in a cohort settings requires incorporating the NxN matrix into the prior distribution. We therefore take a low rank approach to representing the connectome: structural connectomes from 461 subjects (HCP, $n=200$; 1000BRAINS, $n=261$) were parcellated into 20 anatomical atlases (31--294 regions; Table \ref{tab:vcc_parcellations}) and mapped via a linear cross-atlas autoencoder (CrossCoder) into a shared latent space of dimension $n_{\text{lat}} = 16$ (see Section \nameref{sec:cohort_priors}). A cohort-level multivariate normal distribution $\mathcal{N}(\boldsymbol{\mu}, \boldsymbol{\Sigma})$ over this latent space serves as the prior for simulation-based inference (SBI).  We trained a masked autoregressive flow (MAF) on simulated data to learn an amortized posterior $p(\boldsymbol{\theta} \mid \mathbf{x})$, where $\boldsymbol{\theta}$ includes latent connectome codes and dynamical parameters, and $\mathbf{x}$ are subject-specific summary statistics (see Section \ref{sec:amortized_inference}). Once trained, the estimator is shared;
    new subjects are personalized in seconds using only their own data \ref{tab:vep_match_stan}.

  \textbf{Variational cross-coding captures cohort connectivity}
      A prerequisite for cohort-level SBI is that the latent representation reliably identifies individual subjects regardless of atlas. We quantify this with the \emph{confusion rate}: for each subject $s$ and atlas $p$, we encode the connectome to $\mathbf{z}_s^{(p)}$, decode through all other atlases $q \neq p$, re-encode, and measure the fraction of cases where the closest latent neighbor is a different subject.

      For the deterministic CrossCoder, as shown in Figure \ref{fig:vcc_confusion} ($n_{\text{lat}} = 16$), self-decoding (diagonal) identifiability is near-perfect ($> 0.99$). Cross-atlas off-diagonal identifiability averages $0.98$ but ranges down to $0.73$ for certain atlas pairs,
      notably for the 31 region MIST variant,
      confirming a minimum viable number of regions for a useful atlas and that subjects are
      largely separable across atlases: Figure \ref{fig:vcc_latent_weights} shows two subjects' clear and consistent separation in the latent space. The variational mode exhibits a slightly higher rate due to stochastic sampling, but remains well below chance level ($1/231 \approx 0.4\%$). This high self-identification indicates that the cross-coder largely preserves individual identity across granularities, with reduced separability for some atlas pairs (down to 0.73).

      We also evaluated Pearson correlation between original and reconstructed connectomes for each of the 20 atlases. Mean correlation is $r = 0.96$ (range $0.94$--$0.98$) at $n_{\text{lat}} = 16$, with smaller atlases (e.g., 31-region MIST: $r = 0.98$) showing marginally higher fidelity than larger ones (e.g., 294-region JulichBrain: $r = 0.94$) because the latent-to-output ratio is more favorable.


    \textbf{Cohort inference recovers parameters of interest and matches per-subject estimates.}
    To validate that the latent prior enables correct parameter identification, we
    performed cohort-level SBI on synthetic data from the Montbrió-Pazo-Roxin
    (MPR) model (see Section \nameref{sec:montbrio_model}).
      
      \begin{table}[t]
          \centering
          \small
          \begin{tabular}{@{}lcccccc@{}}
          \toprule
          Feature & Training sims & $\dim(\mathbf{x})$ & $s_k$ & $s_D$ & $z_k$ & Coverage \\
          \midrule
          var\_r (baseline) & 4{,}096 & 79 & $0.29 \pm 0.34$ & $0.84 \pm 0.16$ & --- & $53\%$ \\
          var\_rV + cov + eig10 & 8{,}192 & 172 & $0.55 \pm 0.35$ & $0.98 \pm 0.02$ & --- & $57\%$ \\
          \textbf{var\_rV + cov + eig10} & \textbf{16{,}384} & \textbf{172} & $\mathbf{0.57 \pm 0.35}$ & $\mathbf{0.98 \pm 0.03}$ & $\mathbf{1.00}$ & $\mathbf{82.3\%}$ \\
          \bottomrule
          \end{tabular}
          \caption{MPR model posterior diagnostics for cohort-level SBI with different feature extractors and training data budgets. Shrinkage $s = 1 - \sigma_{\text{post}}^2 / \sigma_{\text{prior}}^2$; z-score $z = |\mu_{\text{post}} - \theta_{\text{true}}| / \sigma_{\text{post}}$. Values are mean $\pm$ SD across 231 test subjects. Coverage shows the fraction of true parameters falling within the $90\%$ credible interval for $k$ and $D$ combined.}
          \label{tab:sbi-diagnostics}
      \end{table}

    Of 461 subjects mixed over HCP and 1000BRAINS data, 230 were used for training and 231 held out for test evaluation. Figure \ref{fig:cohort-calibration} shows the distributions of shrinkage and z-score across all 231 test subjects for both parameters. The violin plots confirm that coupling $k$ has moderate but variable identifiability ($s_k = 0.57 \pm 0.35$), with a long tail of poorly identified subjects ($s_k < 0.3$). Noise amplitude $D$ is near-universally well identified ($s_D = 0.98 \pm 0.03$). The z-score distributions are centered near zero for both parameters, confirming that posterior uncertainty bands are well calibrated.
    
    Evaluating different feature configurations across training-data budgets shows that variance-only features plateau at $s_k \approx 0.29$ regardless of data quantity, while FC-enhanced features continue to improve with more data (up to $s_k = 0.57$ at 16{,}384 simulations). Diminishing returns set in above 8{,}192 simulations. Table \ref{tab:sbi-diagnostics} shows the sensitivity of individual summary statistics to $k$ and $D$, confirming that membrane-potential variance ($V$) and covariance eigenvalues are the most informative features for coupling identifiability.  In Figure \ref{fig:shrinkage-zscore}, we demonstrate the final inference performance for
    these features showing good inference on the parameters of interest.
    
    These validations show that the CrossCoder latent space is
    approximately atlas-invariant, largely preserves individual identity, supports accurate simulation-based inference on synthetic data, thus motivating the epilepsy and aging applications below.

\subsection{Characterizing pathology in an epileptic cohort}
  \label{sec:vep_results}

%

\begin{figure}[ht]
    \centering
    \includegraphics[width=\textwidth]{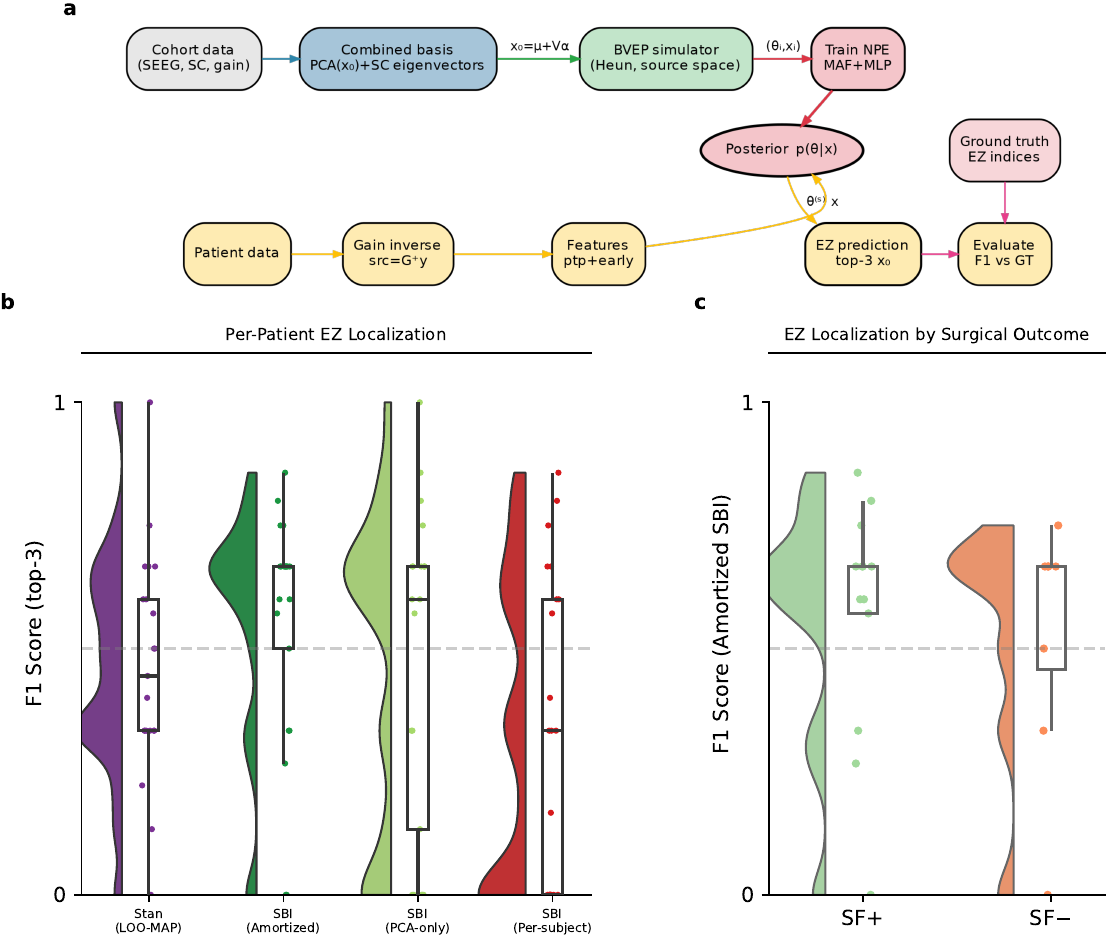}
    \caption{Virtual epilepsy surgery (VEP) results.
    (a)
            \textbf{CAP-VEP amortized inference pipeline.}
            \textbf{Training (top):} Cohort clinical data (SEEG, SC, gain matrices) are used to build the combined basis $V_{x_0}$ (PCA on population $x_0$ patterns + SC eigenvectors). The prior $p(\theta)$ generates parameter vectors, the +c BVEP simulator (Euler integration, source space) generates trajectories $\mathbf{x}(t)$, and features $\mathbf{f}$ (PTP peak-to-peak amplitude, z-scored) are paired with parameters $\theta_i$ to train a neural posterior estimator (NPE-C with MAF and MLP summary network), yielding a single amortized posterior $p(\theta \mid \mathbf{f})$ shared across all patients.
            \textbf{Inference (bottom):} For a new patient, SEEG signals are mapped to source space via a heuristic gain inverse, the same PTP features are extracted, and the trained posterior is conditioned on patient features to sample $\theta^{(s)}$. The top-3 regions by reconstructed median $x_0$ are predicted as the EZ, and evaluated against clinical ground truth (F1 score).
    (b) \textbf{Per-patient F1 scores across four methods.} Amortized SBI and Stan (LOO-MAP) excel on different patients; per-subject SBI consistently underperforms. Seizure-free patients (SF+, \emph{left}) show higher F1 than non-seizure-free (SF$-$) for all methods.
    (c) \textbf{EZ localization accuracy stratified by surgical outcome.} Seizure-free (SF+) patients (validated by curative resection) do not show higher F1 than non-seizure-free (SF$-$) patients.
    }
    \label{fig:vep_results}
\end{figure}

    Drug-resistant epilepsy in 15 million people worldwide drives research
    into better methods for presurgical evaluation. The Virtual Epileptic Patient (VEP) is a personalized virtual brain twin of a patient's seizure propagation pattern (see Section \nameref{sec:generative_models}).  We applied the cohort-amortization principle to build a cohort model of seizure propagation based on clinical evaluation ground truth and structural connectivity and evaluated it against previous state of the art per-patient inference with Stan in a leave-one-out maximum a-posteriori paradigm (LOO-MAP) \citep{Wang2023}. The resulting generative pretrained estimator is efficient and contains no individual patient data. Figure \ref{fig:vep_results}a illustrates the workflow, and the following sections demonstrate its effectiveness.

  \textbf{CAP matches per-subject LOO-MAP.}
    Table~\ref{tab:vep_match_stan} summarizes the main comparison across methods.
    The amortized SBI model achieves mean F1\,=\,0.56 (95\% CI: [0.45,\,0.65]),
    which is comparable to per-patient LOO-MAP (F1\,=\,0.46 [0.37,\,0.56])
    with overlapping confidence intervals (Figure \ref{fig:vep_results}b). This indicates that the cohort-derived informative prior, is sufficient to encode knowledge of typical EZ patterns from the cohort learning the feature-to-posterior mapping from the connectome-driven seizure propagation model that tailors the prior to individual patients.

    \begin{table}[t]
        \centering
        \small
        \begin{tabular}{lcccc}
        \toprule
        Method & F1 Mean [95\% CI] & F1 Median & F1 $>$ 0.5 & Wall Time \\
        \midrule
        Random baseline & 0.02 & 0.00 & 1/21 & --- \\
        Stan (LOO-MAP) & 0.46 [0.37, 0.56] & 0.44 & 8/21 & 2--4 h/patient \\
        \textbf{SBI (Amortized)} & 0.56 [0.45, 0.65] & 0.67 & 15/21 & $\sim$3 min total \\
        SBI (Amortized, PCA-only) & 0.48 [0.33, 0.61] & 0.60 & 13/21 & $\sim$3 min total \\
        SBI (Per-subject, PCA-only) & 0.34 [0.21, 0.47] & 0.33 & 8/21 & 2 min/patient \\
        \bottomrule
        \end{tabular}
        \caption{
          EZ localization across methods. CAP with amortized SBI achieves comparable F1 to 
          state of the art per-patient LOO-MAP with Stan. PCA-only uses the same basis as 
          per-subject SBI to isolate the amortization effect. 
          All SBI evaluations use 500 posterior samples per patient, in less than a minute.
        }
        \label{tab:vep_match_stan}
    \end{table}

    Figure \ref{fig:vep_results}c shows the per-patient F1 scores breakdown.
    Amortized SBI achieves similar F1 scores on seizure-free (SF+) patients (mean 0.57)
    than on non-seizure-free (SF$-$) patients (mean 0.53).

  \textbf{Priors from data drive accuracy.}
    Two primary data-driven priors improved accuracy: principal component analysis
    (PCA) over the clinical evaluation ground truth labels and the eigenvectors of the structural connectivity (SC) together provide an effective low-rank basis for expressing regional pathology $x_0$: the correlation between $\vx_0$ and early source power (seizure onset) across simulated seizures is $\rho = 0.89$ when the SC eigenvector basis is available, versus $\rho = 0.71$ with PCA-only.

    When holding the basis fixed to PCA-only ($K_{\text{SC}} = 0$, 8 basis functions total), the gap between amortized SBI (F1\,=\,0.48) and per-subject SBI (F1\,=\,0.34) is +0.14~F1. This is the pure amortization effect after controlling for representation: the only difference is whether the neural posterior estimator (NPE) was trained on 10{,}000 diverse cohort simulations or 5{,}000 simulations from the patient's own SC. The +0.14 gain suggests that exposure to diverse cohort dynamics may be more informative than deeper sampling of a single patient's connectivity, though the comparison is not matched on simulation count.


    The PCA basis for $x0$ is computed from ground truth EZ labels
    of the same 21 patients used for evaluation, creating a formal risk of data leakage.
    Leave-one-out (LOO) analysis, based on recomputing for each patient the
    8-component PCA basis from the remaining 20 patients and measuring subspace alignment,
    showed mean alignment of $0.97 \pm 0.04$ (minimum $0.87$ for patient id027\_sj).
    This high stability confirms the 8-component basis is overdetermined by the
    21-patient cohort: removing any single patient shifts the principal subspace
    by less than 3\%, confirming that individual EZ patterns do not dominate the
    learned spatial modes.

\subsection{Structure-function coupling in an aging cohort}
    \label{sec:aging_detailed_results}

    \begin{figure*}[p]
      \centering
      \includegraphics[width=\textwidth]{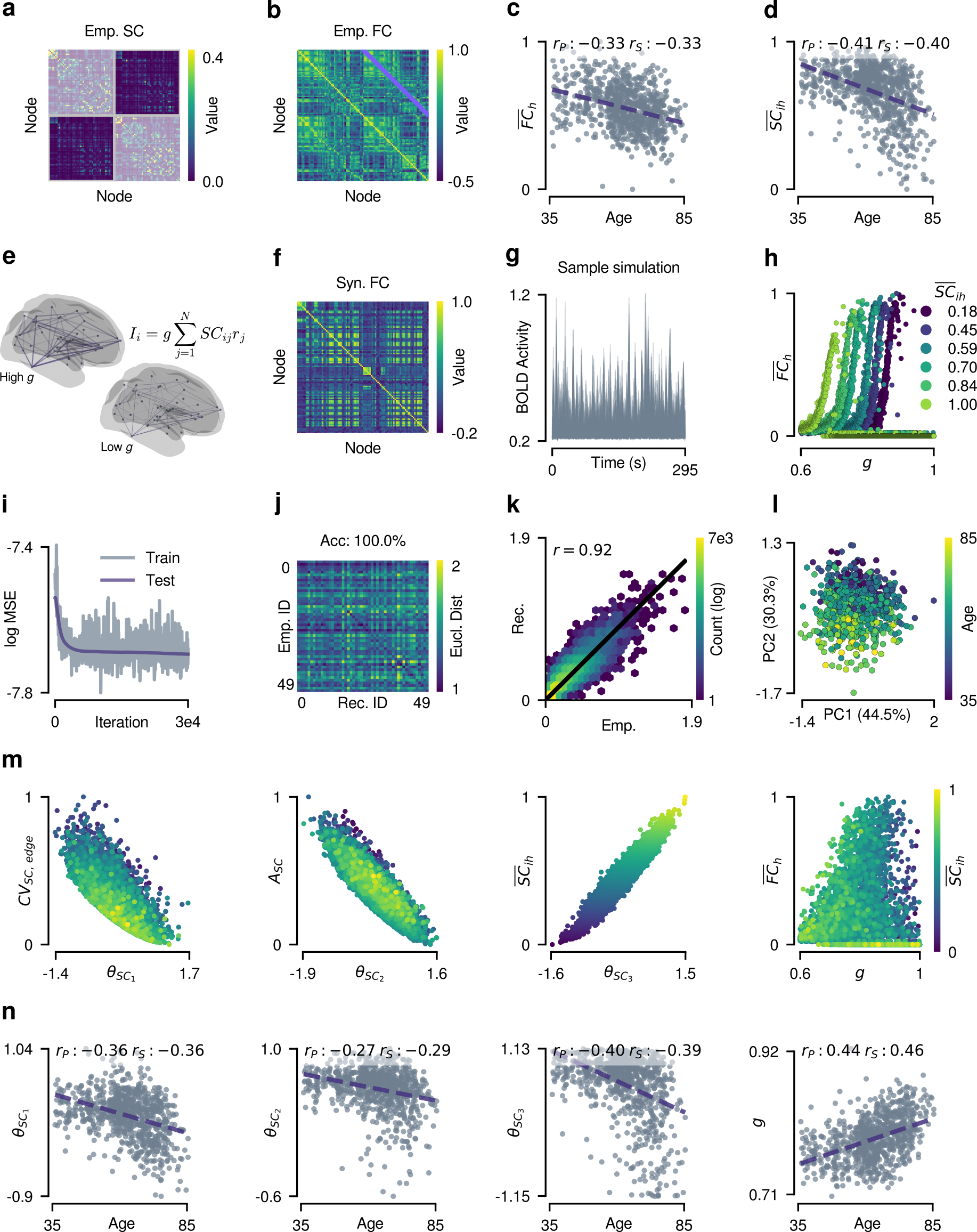}
      \caption{\textbf{Cohort-amortized inference for the 1000BRAINS aging cohort.}
      (a) Empirical structural connectivity matrix. (b) Empirical functional connectivity matrix, with homotopic connections highlighted in light purple. (c) Homotopic functional connectivity versus age. (d) Interhemispheric structural connectivity versus age. (e) Schematic representation of the mechanistic role of global coupling parameter $g$. (f-g) Synthetic functional connectivity matrix and its respective BOLD time-series. (h) Feature-to-parameter ($\overline{FC}_{\mathrm{h}}$ to $g$) curves obtained from per-subject simulations. (i) CrossCoder training loss over 30,000 iterations. (j) Leave-one-out subject identifiability confusion matrix. (k) Latent space reconstruction fidelity (Euclidean distance). (l) PCA projection of the 3D CrossCoder latent space, colored by normalized subject age (purple: younger; yellow: older). (m) Joint distribution of parameters and features derived from $N=10,000$ synthetic simulations. (n) Inferred posterior modes for structural latent parameters ($\theta_{\mathrm{SC}_1}, \theta_{\mathrm{SC}_2}, \theta_{\mathrm{SC}_3}$) and global coupling $g$ as a function of chronological age.}
      \label{fig:fig_mpr_main}
    \end{figure*}
    
    \begin{figure*}[ht]
      \centering
      \includegraphics[width=\textwidth]{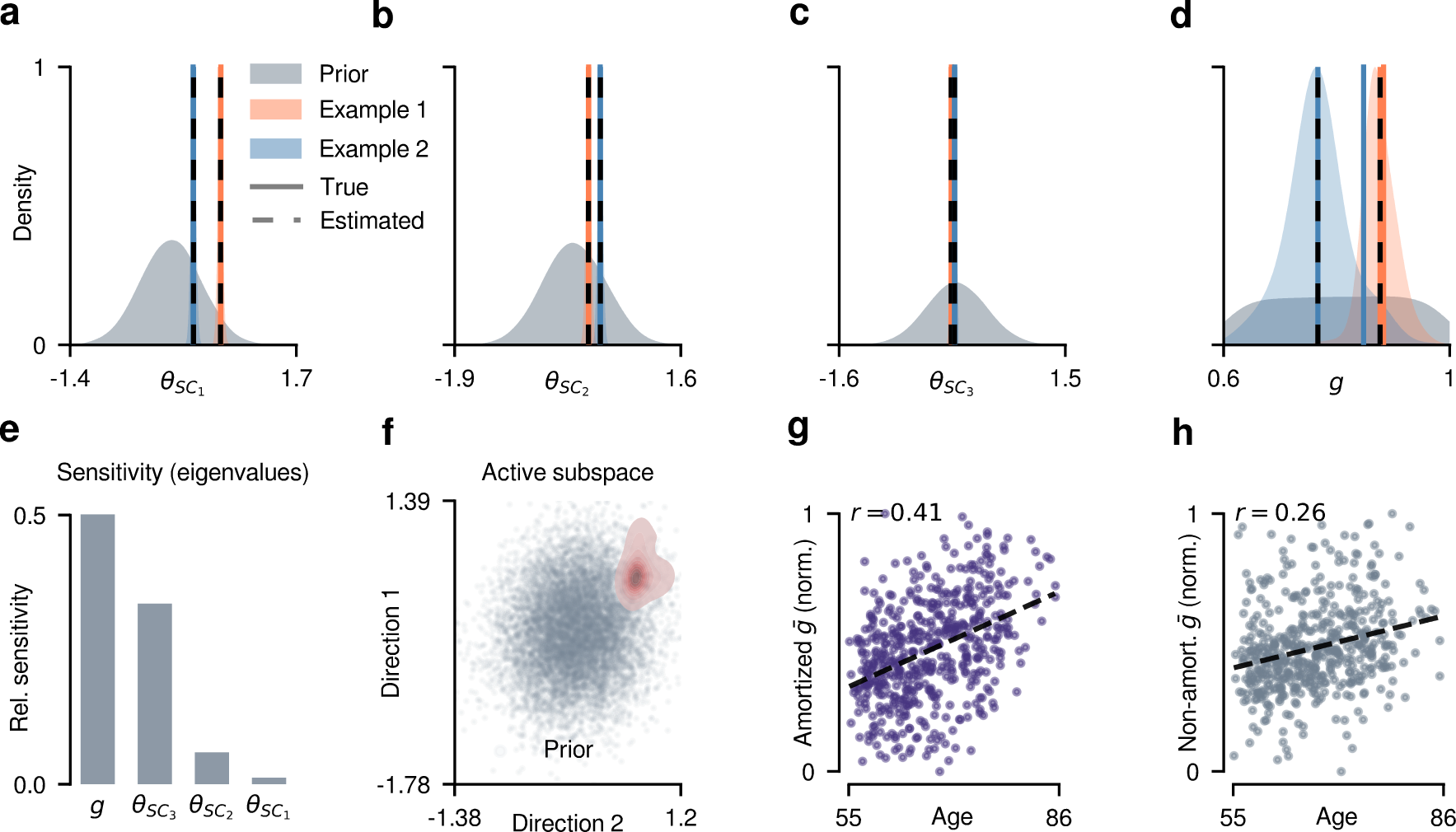}
      \caption{\textbf{Robustness and validation of the cohort-amortized pipeline.} \textbf{(a--d)} Parameter recovery for a pair of held-out simulations (Example~1, orange; Example~2, blue). For each inferred parameters ($\theta_{SC_1}$, $\theta_{SC_2}$, $\theta_{SC_3}$, and $g$), distributions show prior (gray) and amortized posterior (colored), solid lines mark true values, and dashed lines mark the posterior means (from 1{,}000 samples). In all cases the posterior shrinks substantially relative to the prior, with true values consistently within the high-density regions, confirming successful recovery. The wider posterior for $g$ reflects its inherently lower signal-to-noise ratio compared to the structural latent dimensions, consistent with the sensitivity ranking. \textbf{(e)} Parameter sensitivity, quantified by the eigenvalue-weighted contribution to the leading active subspace directions. Global coupling $g$ dominates, followed by $\theta_{SC_3}$, confirming that the inference is primarily driven by parameters with the largest impact on the summary statistics. \textbf{(f)} Two-dimensional projection of prior samples (gray) and a representative posterior (red, kernel-density contour) onto the two leading active subspace directions. The concentrated posterior contrasts with the diffuse prior, illustrating that the learned estimator effectively narrows the plausible values to a small submanifold of prior support. \textbf{(g)} Amortized posterior mean $\tilde{g}$ versus chronological age  (1000BRAINS sub-cohort, $n = 502$, Pearson $r = 0.41$). \textbf{(h)} Non-amortized (per-subject) $\tilde{g}$ versus age  ($r = 0.26$). The stronger correlation confirms that the amortized inference yields superior age-related signal while reducing total inference time by orders of magnitude.}
      \label{fig:fig_supp_robustness}
    \end{figure*}

    Beyond personalized inference, cohort-amortized personalization maps cohort-level distributions (i.e. means and covariances represented by multivariate normal distributions), all while minimizing the computational cost of probabilistic inference by orders of magnitude. While the exact number of simulation budget depends on the diversity and heterogeneity of summary statistics across the cohorts, it can be typically reduced from from $n_{sim} = n_{sim_{subject}} * n_{subject}$ in the non-amortized case, to $n_{sim} = n_{sim_{subject}} * k$ in amortized implementations, where $k$ is a small scaling factor ensuring adequate sampling from the joint prior distribution of the generative model.
    
    We therefore constructed a CAP pipeline around the 1000BRAINS aging cohort as a large-scale population benchmark to systematically evaluate the network's propensity for faithful, efficient, and reproducible probabilistic inference from high-dimensional datasets. Specifically, we aimed to invert the global coupling parameter $g$, which previous non-amortized inference work demonstrated as a potential identifier of healthy aging and its associated decline \citep{Lavanga2023}.
    
    This pipeline consisted of training the autoencoder on structural connectivity (facilitating a low-dimensional SC parameterization), which was leveraged to conduct the amortized inference training built upon $10{,}000$ simulation budget. Finally, we inferred the global coupling parameter $g$ corresponding to the chosen empirical observations (i.e. summary statistics) spanning $832$ subjects in the age range of $[35, 85]$, the results of which are organized around the panels shown in Figure~\ref{fig:fig_mpr_main}, with additional validation in Figure~\ref{fig:fig_supp_robustness}.
    
\textbf{Macroscale biomarkers of aging.}
    Previous research has corroborated corpus callosum (CC) atrophy \citep{Sullivan2010} during age-related decline. As such, a primary observable for characterizing empirical structural connectomes (Figure~\ref{fig:fig_mpr_main}a) is the prominence of interhemispheric connections (off-diagonal quadrants). CC atrophy reportedly leads to downstream functional dedifferentiation \citep{Cabeza2002, Lavanga2023}, which --in terms of macroscale observables-- can be best characterized by changes in the overall prominence of homotopic region pairs; this reflects a breakdown in the strong functional coupling between typically observed between these regions despite only moderate direct structural connectivity (Figure~\ref{fig:fig_mpr_main}b). Crucially, both of these macroscale properties degrade systematically with age: mean homotopic FC ($\overline{FC}_h$) declines significantly across the cohort (Figure~\ref{fig:fig_mpr_main}c; $r_P = -0.33$, $r_S = -0.33$, both $p < 10^{-30}$), as does mean interhemispheric SC ($\overline{SC}_{\mathrm{ih}}$; Figure~\ref{fig:fig_mpr_main}d; $r_P = -0.41$, $r_S = -0.40$, both $p < 10^{-45}$). These measures capture partially dissociable aspects of the aging process, making them complementary summary statistics for the amortized inference pipeline.
    
\textbf{Prior distribution and simulation framework.}
    Figure~\ref{fig:fig_mpr_main}e illustrates a schematic  of previously-established in-silico characterization of healthy aging through the global coupling parameter $g$:  the coupling term $I_i = g \sum_j SC_{ij} r_j$ dynamically mediates inter-regional communication, whereby the global scaling factor $g$  determines the magnitude by which structural connectomes $SC_{ij}$ influence regional firing rates. High values of $g$ amplify inherently weak structural connections, facilitating distributed functional integration even in the face of degraded white matter pathways (as depicted in the bottom brain schematic simulating older adults); conversely, low values of $g$ tightly restrict functional coupling solely to the strongest structural connections (Figure~\ref{fig:fig_mpr_main}e, `High g'
    schematic of younger adults with intact commissural pathways).
    
    The prior over $g$ was therefore chosen as a uniform distribution over the interval $[0.6, 1.0]$, such that for each 'synthetic' subject (i.e. simulations with highly similar connectome scaling), simulations would cover an extensive range of -weak to strong- structure-function correspondence. This specific range was calibrated in order to generate plausible BOLD dynamics: given the scale of input structural connectomes, values of $g < 0.6$ yield quiescent dynamics that conflict with empirical rs-fMRI dynamics, whereas $g > 1.0$ artificially drives the system into pathological synchronization. This flat prior distribution is thus informative at the cohort scale, but uninformative at the individual scale, reflecting our agnostic stance and permitting the resulting individual posterior estimates to be uniquely determined by each subject's empirical features. A representative trace of simulated BOLD activity, and the functional connectome derived thereof showcase spontaneous fluctuations exhibiting coordination patterns consistent with empirical resting-state fMRI (Figure~\ref{fig:fig_mpr_main}
    f-g). Since the previous non-amortized attempt was conducted using a slightly older preprocessing pipeline, we attempted to establish a non-amortized baseline using the abovementioned prior configuration. As such, we computed personalized feature-to-parameter curves obtained through per-subject forward simulations across the viable range of $g$. Said personalized simulations were attained using a subset of empirical structural connectomes chosen to be equidistant along a normalized, strictly increasing trajectory of connectome scaling measures ($\overline{SC}$ 10th percentile, $\overline{SC}_{\mathrm{ih}}$, and largest eigenvalue $\lambda_{\max}$ (Figure~\ref{fig:fig_mpr_main}h).

\textbf{Crosscoder training and validation.}
    We trained the deterministic CrossCoder for 30{,}000 iterations (Figure~\ref{fig:fig_mpr_main}i), to compress connectomes from the Schaefer 17-network atlas (100-parcels), into a 3-dimensional latent space while preserving subject-specific characteristics. The overlap across training and testing traces alludes to the absence of overfitting despite the high compression ratio, in which 4{,}950 SC edges were compressed into 3 latent dimensions, representing a 1{,}650-fold reduction. Subject identifiability was assessed via leave-one-out cross-reconstruction (Figure~\ref{fig:fig_mpr_main}j): We computed Euclidean distances separating each subject's latent-state representations from all others, and subsequently pinpointed the absolute nearest neighbor. The visually prominent diagonal in the confusion matrix confirms the robustness of per-subject identifiability. 
    The 100\% nearest-reconstruction accuracy on a 100-subject subset indicates that individual SC structure is well separated in the latent space for the tested subjects. Retaining this degree of identifiability is instrumental for conducting inference on the cohort scale. Specifically, multiple subjects collapsing onto the exact same latent coordinates translates to a degenerate mapping between macroscale observables and latent state parameters, and a subsequent failure in posterior inference. Additionally, the fidelity of reconstruction is quantified in Figure~\ref{fig:fig_mpr_main}k, which illustrates a density scatter plot (i.e. hexbin) of empirical vs. reconstructed individual SC edge weights, pooled across all subjects and all edges (4{,}950 edges per subject). The distribution of points (log count) along the diagonal ($r = 0.91$) validates the decoding accuracy of individual edge weights across the full range of connectivity values, without a systematic bias toward over/underestimation at either end of the distribution. As revealed by PCA of the latent space (Figure~\ref{fig:fig_mpr_main}l), inter-subject variability of structural connectomes is well captured by a continuous gradient navigating through the latent space. Within this representation, each point denotes one subject, color-coded by normalized age (purple: younger; yellow: older). The first two principal components (PC1 and PC2) account for 44.5\% and 30.3\% of total latent variance, respectively, with chronological age primarily aligning with the second principal axis.

\textbf{Parameter-feature relationships and posterior identifiability.}
    The quantitative and qualitative characteristics of feature-to-parameter mappings directly influence the invertible functions learned by the NPE. As such, amortized inference should include not only data features which best represent changes to neural dynamics, but also those which best capture personalized characteristics (e.g. features characterizing different aspects of structural connectomes). Figure~\ref{fig:fig_mpr_main}m precisely visualizes such mappings across the $N = 15{,}000$ training simulations. Here, each distinct point represents a single simulation, color-coded by $\overline{SC}_{\mathrm{ih}}$, which is chosen to represent the overall approximate scale of interhemispheric connections:
    
    As shown, the first latent dimension $\theta_{\mathrm{SC}_1}$ is negatively associated with the coefficient of variation of SC edge weights ($\mathrm{CV}_{\mathrm{SC,edge}}$), such that connectomes exhibiting high $\theta_{\mathrm{SC}_1}$ possess low SC heterogeneity (signifying highly uniform edge weight distributions), whereas those exhibiting low $\theta_{\mathrm{SC}_1}$ display high SC heterogeneity (reflecting a hub-rich connectivity defined by strong core edges and distinctly weak peripheral edges). This dimension may therefore be used as a proxy to the degree of scale-freeness of subjects' connectomes.
    
    Similarly, the second latent dimension $\theta_{\mathrm{SC}_2}$ correlates negatively with the SC asymmetry index $A_{\mathrm{SC}}$, in that subjects with high $\theta_{\mathrm{SC}_2}$ present congruent intrahemispheric wiring between the two hemispheres, whereas low $\theta_{\mathrm{SC}_2}$ corresponds to a greater structural divergence between left and right intrahemispheric subnetworks.
    
    The third and final structural latent dimension $\theta_{\mathrm{SC}_3}$ illustrates a near-linear correspondence with mean interhemispheric structural connectivity $\overline{SC}_{\mathrm{ih}}$, which constitutes one pronounced characteristic of brain aging (Figure~\ref{fig:fig_mpr_main}d). This dimension can thus be interpreted as a privacy-preserving proxy for an anatomical substrate associated with age-related structural disconnection. The robustness of this near-linear mapping confirms that $\theta_{\mathrm{SC}_3}$ is identifiable directly from empirical data; given a subject's $\overline{SC}_{\mathrm{ih}}$, the posterior over $\theta_{\mathrm{SC}_3}$ will concentrate tightly around the corresponding latent value. Note that $A_{\mathrm{SC}}$ and $\overline{SC}_{\mathrm{ih}}$ offer complementary and structurally non-overlapping representations of bilateral organization. Including both therefore ensures that the summary statistics encode distinct aspects of hemispheric SC that are jointly relevant to the inference of $g$.
    
    As portrayed through the equations laid out in (Figure~\ref{fig:fig_mpr_main}e), multiplying the global coupling parameter $g$ by each subject's connectivity matrix determines the personalized working points within the context of BOLD feature-to-parameter curves. Sampled parameters from the learned latent multivariate normal distributions to attain reconstructed `synthetic' subjects, with each point in the feature-parameter scatters formally representing a single instance of such a subject. However, due to the fundamental role of the global and interhemispheric connections in modulating the personalized working point for each subject, the more practical perspective is to view connectomes/simulations with a similar degree of degradation as an aggregate synthetic subject (Figure~\ref{fig:fig_mpr_main}m, last panel), which facilitates one in drawing a parallel with per-subject empirical simulations illustrated in (Figure~\ref{fig:fig_mpr_main}h).
    
    \textbf{Empirical posterior estimates and brain health trajectories.}
    The density estimator was then trained on the set of joint prior samples, and synthetic data features computed thereof. Subsequent to this training, empirical observations (i.e. summary statistics) were used to estimate parameters pertaining to latent structural states and the global coupling parameter. Figure~\ref{fig:fig_mpr_main}n presents the per-subject inferred posterior means for the four aforementioned parameters as a function of subject age. In line with their corresponding empirical features, the three structural latent parameters demonstrate a significant decline with advancing age ($r_P = -0.36$, $r_S = -0.36$ for $\theta_{SC_{1}}$; $r_P = -0.27$, $r_S = -0.29$ for $\theta_{SC_{2}}$; and $r_P = -0.40$, $r_S = -0.39$ for $\theta_{SC_{3}}$), corroborating the role of each parameter in encoding unique but integral aspects of connectomes, namely $\mathrm{CV}_{\mathrm{SC,edge}}$, $A_{\mathrm{SC}}$, and $\overline{SC}_{\mathrm{ih}}$ \citep{Betzel2014}. Most crucially, global coupling $g$ increases significantly with age ($r_P = 0.44$, $r_S = 0.46$, both $p < 10^{-20}$), opposing the observed trend in inferred structural latents, which is consistent with previously reported non-amortized inference results on the same cohort (Figure~\ref{fig:fig_mpr_main}n) \citep{Lavanga2023}. Robustness checks including synthetic parameter recovery, sensitivity analysis via active subspaces, and a direct comparison between amortized and non-amortized posterior estimates for $g$ are reported in Figure~\ref{fig:fig_supp_robustness}. This finding further demonstrates the preservation of population-level trends in our amortized pipeline, while eliminating the computational overhead  typically associated with per-subject inference.

\section{Discussion}\label{sec:discussion}


  Across both the synthetic validation, clinical epilepsy and aging cohorts, the results demonstrate that CAP achieves practical deployability with accuracy comparable to per-subject inference: the CrossCoder latent space preserves individual subject identity across 20 heterogeneous atlases and supports high-fidelity reconstruction, enabling atlas-agnostic deployment across sites. This shared latent prior drives accurate simulation-based inference, with feature selection remaining the primary determinant of parameter identifiability. In the clinical setting, amortized inference reduces wall time from 2–4 hours per patient to approximately three minutes total while matching or exceeding per-subject accuracy (Table \ref{tab:vep_match_stan}) due to cohort data enabling better
  model priors, and incorporating structural connectivity eigenvectors provides meaningful gains in focal epileptogenic zone localization, the estimation of structure-function correspondence in an aging population. Heterogeneous cohort training outperformed per-subject training in our comparison ($+0.14$ F1, $n=21$, PCA-only basis), suggesting that exposure to diverse population dynamics may be more informative than additional single-patient simulation, though the two settings also differed in simulation budget, thereby establishing CAP as a computationally viable, privacy-preserving platform for both research and clinical translation.

  CAP's primary constraint is that not all patient-specific data are amenable to compact prior representations: in the epilepsy use case, implantation-specific SEEG gain matrices and sensor geometries required a source-space projection workaround that may not generalize to other modalities.
  The clinical evaluation is further limited by the 21-patient cohort, which yields wide confidence intervals and no statistically significant paired differences between methods; the fixed top-3 EZ selection metric under-recalls patients with diffuse epileptogenic zones; and hyperparameters were tuned on the same small cohort without independent validation, likely inflating reported performance.
  Furthermore, our capacity to estimate neuromodulation outcomes and brain health trajectories in the aging population is currently limited by a lack of empirical ground truth. Validating these stimulation counterfactuals will ultimately require integrating longitudinal and interventional data rather than relying exclusively on cross-sectional models.
  Finally, the CrossCoder is restricted to linear encoder–decoder pairs, which showed less overfitting than nonlinear alternatives in cross-validation but may limit reconstruction fidelity for pathological connectomes, while the variational mode trades slightly higher confusion rates for uncertainty quantification, making the deterministic mode preferable when reproducibility is paramount.

  Normative modeling (NM) charts subject-level deviations from population trajectories, with federated strategies now enabling cross-site harmonization and deployment without data pooling \citep{Marquand2016, Kia2022}. NM is descriptive, flagging which features deviate from the expected ranges, whereas CAP is mechanistic, identifying the generative parameters that explain an individual's data. The approaches are complementary: normative deviations identify atypical individuals, while CAP provides mechanistic explanations.

  Federated learning (FL) iteratively aggregates locally computed gradients so raw data never leave the institution \citep{McMahan2017}, yet practical limitations persist: FedAvg fails for very small centers and repeated communication rounds introduce substantial latency \citep{Denissen2026}. Synthetic-data generation combined with federated networks offers a complementary privacy-preserving route \citep{Bajwa2025}. CAP is naturally compatible with both: because the amortized estimator is trained on simulated rather than raw observations, model sharing replaces data sharing, and the one-time training cost avoids the iterative communication overhead of standard FL. Where multi-site collaboration is possible, CAP can be combined with federated training of priors and estimator, but its core advantage is that a single pretrained network can be shared and applied locally without any ongoing data transmission.

  Foundation models and autoencoder-based approaches, such as Krakencoder, 
  map connectomes into reusable latent spaces across modalities and atlases \citep{Moor2023, Jamison2025}, yet these representations are not directly interpretable: latent dimensions must still be correlated with clinical or biological variables post hoc. CAP closes this gap by conditioning the mechanistic generative model directly on latent representations as priors, inferring posterior distributions over biophysical parameters rather than remaining at the level of representation. Because CAP is built on coupled neural mass models, inferred parameters describe changes in coupling, excitability, or noise, and the model can predict how dynamics would change under stimulation or virtual lesions. In this sense, representation models provide the latent connectome space, and CAP provides the mechanistic personalization layer that turns shared representations and local empirical data into individualized, interpretable virtual brain twin predictions \citep{Wang2025, Hashemi2025}.

  CAP enables deployment across clinical, cloud, and edge computing environments. The trained inference network is compact (approximately one million parameters) and produces personalized posteriors in under five seconds on a standard GPU, with cloud deployment via the EBRAINS infrastructure providing sub-minute turnaround for clinical workflows \citep{Schirner2022,Baldy2025}. Critically, because the network encodes cohort-level statistics rather than individual training data, it is intended to be shareable under common data-protection frameworks, subject to institutional review.
  
  More broadly, the shared artefact is a \textit{mechanistic surrogate}: the combination of a learned cohort distribution and an explicit generative simulator may support synthetic-cohort generation, virtual lesions, stimulation counterfactuals, and parameter sensitivity analyses without access to raw individual data, within the limits of the generative model \citep{SanzLeon2013, Alstott2009, Dollomaja2025, Constantine2015}.
  Precedents for this kind of work already exist. Virtual brain twins building on the
  the Virtual Brain platform have long supported in-silico perturbation experiments \citep{SanzLeon2013}; synthetic epilepsy cohorts have been released on EBRAINS for external benchmarking~\citep{Dollomaja2025}; and whole-brain models have been used to study age-related dedifferentiation and network
  resilience~\citep{Gatica2022, Lavanga2023, Achard2006}. CAP does not create
  these scientific possibilities; it may remove governance and computational
  barriers that currently limit their replication across institutions.
  Within the limits of the assumptions of the generative model, synthetic-access research may provide a middle path between fully open data, with its privacy risks, and data enclaves, which do not scale. In this perspective, personal data remain local, and the pretrained model travels, allowing for wider adoption of the models in more diverse settings, paving the way to address new research questions \citep{Eke2022, Dollomaja2025, Vaden2020}.

\section{Methods}

\subsection{Data cohorts}
\label{sec:data_cohorts}

  \subsubsection{Epilepsy cohort}
  \label{sec:epilepsy_cohort}
    We used the data from patients who underwent a standard presurgical protocol at La Timone Hospital in Marseille.
    Informed written consent was obtained in compliance with the ethical requirements of the Declaration of Helsinki and the study protocol was approved by the local Ethics Committee (Comité de Protection des Personnes sud Méditerranée 1). Each patient underwent a comprehensive presurgical assessment, which included medical history, neuropsychological assessment, neurological examination, fluorodeoxyglucose positron emission tomography, high-resolution 3T MRI, long-term scalp-EEG and invasive SEEG recordings. They received non-invasive T1-weighted imaging (MPRAGE sequence, repetition time = $1.9\ s$, echo time = $2.19\ ms$, voxel size $1.0\times1.0\times1.0\ mm^3$) and diffusion-weighted MRI images (with an angular gradient set of 64 directions, repetition time = $10.7\ s$, echo time = $95\ ms$, voxel size $1.95\times 1.95\times 2.0\ mm^3$, b-weighting of $1000\ s/mm^2$.) The images were acquired on a Siemens Magnetom Verio 3T MR-scanner. The patients had invasive SEEG recordings obtained by implanting multiple depth electrodes, each containing $10-18$ contacts $2\ mm$ long and separated by $1.5$ or $5\ mm$ gaps. The SEEG signals were acquired on a 128 channel Deltamed/Natus system. After the electrode implantation, a cranial CT scan was performed to obtain the location of the implanted electrodes.
    Ground-truth EZ labels were available for 23 patients based on surgical outcomes; of these, 2 patients were excluded from evaluation due to missing SEEG information, structural connectivity or incomplete electrode coverage, 
    leaving 21 patients (13 seizure-free, SF+; 8 non-seizure-free, SF$-$).

    To construct the individual brain network models, we first preprocessed the T1-MRI and diffusion-weighted MRI data. Volumetric segmentation and cortical surface reconstruction were from the patient-specific T1-MRI data using the recon-all pipeline of the FreeSurfer software package (\url{http://surfer.nmr.mgh.harvard.edu}). The cortical surface was parcellated according to the VEP atlas \citep{Wang2021},  the code for which is available at \url{https://github.com/HuifangWang/VEP_atlas_shared.git}. The reasons for choosing the VEP atlas are as follows: 1) The VEP atlas is specifically designed for epileptology, incorporating anatomical and functional features of each brain structure, particularly in relation to the EZN and surgical applications. 2) The geometric features and sizes of the brain regions are well suited for both clinical applications and modeling, including model inversion. 3) Brain regions can be automatically labeled and personalized from T1-MRI scans using geometric and neuroanatomical information. 4) The VEP atlas has been clinically evaluated in a retrospective study including 53 patients, and is currently being assessed in another prospective trial with 356 patients, ensuring its clinical suitability and reliability \citep{Wang2023,Jirsa2023,Makhalova2022}.
    A brief summary of the steps to obtain the VEP atlas goes as follows. T1-weighted images are processed using FreeSurfer to remove non-brain tissue \citep{Segonne2004}, segment subcortical gray matter structures \citep{Fischl2004}, normalize intensity \citep{Sled1998}, and generate cortical surfaces \citep{Segonne2007}. These surfaces are inflated, registered to a spherical template, and corrected for topology \citep{Fischl1999,Fischl2004}. The cortex is then subdivided into regions based on gyral and sulcal structures \citep{Destrieux2010}, forming the basis for constructing the VEP atlas. This construction involves splitting, merging, and renaming operations to both cortical and subcortical regions. Cortical regions are divided based on the triangulated surface mesh, while subcortical regions are split directly on voxels. Nonlinear splits are applied in specific areas with high curvature, such as the callosal sulcus, whereas the superior frontal gyrus is split using a combination of linear and specialized methods based on cortical surface geometry \citep{Wang2021}.

    We used the MRtrix software package to process the diffusion-weighted MRI \citep{Tournier2019}, employing the iterative algorithm described in \citep{Tournier2012} to estimate the response functions and subsequently using constrained spherical deconvolution \citep{Tournier2007} to derive the fiber orientation distribution functions. The iFOD2 algorithm \citep{Tournier2010} was used to sample 15 million tracts. The structural connectome was constructed by assigning and counting the streamlines to and from each VEP brain region. The diagonal entries of the connectome matrix were set to 0 to exclude self-connections within areas and the matrix was normalized so that the maximum value was equal to one. We obtained the location of the SEEG contacts from post-implantation CT scans using GARDEL (Graphical User Interface for Automatic Registration and Depth Electrode Localization), which is part of the EpiTools software package \citep{MedinaVillalon2018}. Then we coregistered the contact positions from the CT scan space to the T1-MRI scan space for each patient.
    Atlas region centers and contact positions were used to compute a
    gain matrix $G \in \mathbb{R}^{n_s \times 162}$ (with $n_s$ varying by patient) that maps source-level activity onto the electrode sensors through a linear forward model $y = G \cdot x + \epsilon$ under a dipole in a homogeneous infinite medium approximation \citep{Sarvas1987}.

  \subsubsection{1000BRAINS cohort}
  \label{sec:1000brains_cohort}

    The 1000BRAINS project is a population-wide study of healthy aging that characterizes the variability of brain structure and function in relation to behavioral factors \citep{Caspers2014}. Full details of recruitment, eligibility, MR acquisition (3T Siemens Tim-TRIO, 32-channel head coil),
    and the structural, diffusion, and resting-state protocols are given in the original cohort description \citep{Caspers2014}. Diffusion processing (multi-shell HARDI; multi-tissue CSD and iFOD2 probabilistic tractography in MRtrix3 with SIFT2 streamline weighting) and the construction of connectivity matrices
    over the 100-region, 17-network Schaefer 2018 atlas \citep{Schaefer2018,Tournier2019,Smith2015SIFT2} follow the released pipeline. From this release we only retained the first visit of subjects older than 35 years for whom both diffusion and resting-state functional MRI scans of acceptable quality were available, yielding the final sample used here ($n = 832$, $n_{\mathrm{Female}} = 368$). Resting-state fMRI (gradient-echo EPI, TR\,=\,2200\,ms, 300 volumes, $\approx$11\,min) was preprocessed with fMRIPrep v24.1.1 \citep{Esteban2019} and denoised using the Nilearn interface
    to the fMRIPrep confounds following the \emph{simple} (V4) strategy of
    \citet{Ciric2017} (cosine-basis high-pass filtering, 24-parameter motion regression, and mean WM/CSF regression, without global signal regression or scrubbing). Regional mean time series were extracted over the 100 Schaefer parcels \citep{Schaefer2018}.

    We opted for minimal post-processing of structural connectomes and BOLD time-series. For structural connectomes, we chose raw SIFT2-weighted matrices, where each subject's connectivity matrix ($n_{nodes}=100$) was loaded and aggregated into a cohort array. This array was subsequently normalized by its maximum value, such that all edge weights lie in $[0, 1]$. The BOLD time-series were used as--is for the computation of functional connectomes.

\subsection{Cohort priors}
\label{sec:cohort_priors}

  \subsubsection{CrossCoder architecture}
  \label{sec:crosscoder_architecture}

\begin{figure}[t]
    \centering
    \begin{tikzpicture}[
        node distance=1.2cm and 1.5cm,
        box/.style={draw, rounded corners, fill=lightblue, align=center, minimum width=2.2cm, minimum height=0.9cm, font=\small},
        latent/.style={draw, rounded corners, fill=darkblue!15, align=center, minimum width=3cm, minimum height=1.2cm, font=\small},
        arrow/.style={->, thick, >=stealth},
        label/.style={font=\footnotesize\itshape}
    ]

    \node[box] (p1) at (0,4) {31-region\\MIST};
    \node[box] (p2) at (0,2.8) {79-region\\Shen};
    \node[box] (p3) at (0,1.6) {150-region\\Destrieux};
    \node[box] (pdots) at (0,0.6) {$\vdots$};
    \node[box] (p20) at (0,-0.4) {294-region\\JulichBrain};

    \node[box, fill=greenaccent!20] (enc1) at (3,4) {Enc$_1$};
    \node[box, fill=greenaccent!20] (enc2) at (3,2.8) {Enc$_2$};
    \node[box, fill=greenaccent!20] (enc3) at (3,1.6) {Enc$_3$};
    \node[box, fill=greenaccent!20] (enc20) at (3,-0.4) {Enc$_{20}$};

    \node[latent] (latent) at (6.5,1.8) {$\mathbf{z} \in \mathbb{R}^{n_{\text{lat}}}$\\$n_{\text{lat}} \in \{8,16,32\}$};

    \node[box, fill=accent!15] (mvn) at (6.5,-0.8) {Cohort MVN\\$\mathcal{N}(\boldsymbol{\mu}, \boldsymbol{\Sigma})$};

    \node[box, fill=greenaccent!20] (dec1) at (10,4) {Dec$_1$};
    \node[box, fill=greenaccent!20] (dec2) at (10,2.8) {Dec$_2$};
    \node[box, fill=greenaccent!20] (dec3) at (10,1.6) {Dec$_3$};
    \node[box, fill=greenaccent!20] (dec20) at (10,-0.4) {Dec$_{20}$};

    \node[box] (o1) at (13,4) {$\hat{\mathbf{W}}^{(1)}$};
    \node[box] (o2) at (13,2.8) {$\hat{\mathbf{W}}^{(2)}$};
    \node[box] (o3) at (13,1.6) {$\hat{\mathbf{W}}^{(3)}$};
    \node[box] (o20) at (13,-0.4) {$\hat{\mathbf{W}}^{(20)}$};

    \draw[arrow] (p1) -- (enc1); \draw[arrow] (enc1) -- (latent);
    \draw[arrow] (p2) -- (enc2); \draw[arrow] (enc2) -- (latent);
    \draw[arrow] (p3) -- (enc3); \draw[arrow] (enc3) -- (latent);
    \draw[arrow] (p20) -- (enc20); \draw[arrow] (enc20) -- (latent);

    \draw[arrow] (latent) -- (dec1); \draw[arrow] (dec1) -- (o1);
    \draw[arrow] (latent) -- (dec2); \draw[arrow] (dec2) -- (o2);
    \draw[arrow] (latent) -- (dec3); \draw[arrow] (dec3) -- (o3);
    \draw[arrow] (latent) -- (dec20); \draw[arrow] (dec20) -- (o20);

    \draw[arrow, dashed, color=accent] (latent) -- (mvn);

    \node[label, above=0.1cm of p1] {Input connectomes};
    \node[label, above=0.1cm of o1] {Reconstructed};
    \node[label, below=0.1cm of mvn] {SBI prior};

    \end{tikzpicture}
    \caption{CrossCoder architecture. Twenty atlases with different dimensionalities ($d_p = n_p(n_p-1)/2$) are mapped through per-view linear encoders into a shared latent space $\mathbf{z} \in \mathbb{R}^{n_{\text{lat}}}$. A cohort-level multivariate normal (MVN) denoted by $\mathcal{N}(\boldsymbol{\mu}, \boldsymbol{\Sigma})$ over the latent space serves as the prior for simulation-based inference. Decoders map latent codes back to each atlas, enabling cross-atlas connectome synthesis.}
    \label{fig:vcc_architecture}
    \end{figure}
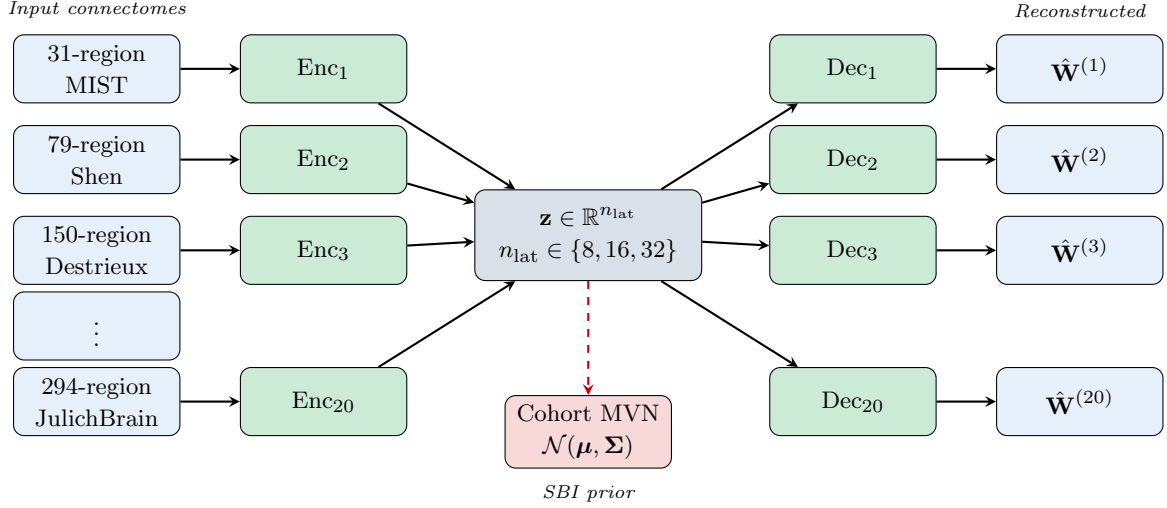

    \textbf{CrossCoder architecture}\label{sec:architecture}
    Structural connectomes are derived from diffusion MRI tractography and parcellated according to anatomical atlases. Different studies employ different atlases---ranging from 31 to 294 regions---which complicates cohort-level analysis because connectome dimensionality varies across datasets. We address this with a \emph{CrossCoder} (see Figure \ref{fig:vcc_architecture}), a linear autoencoder that maps connectomes from any atlas into a shared latent space of fixed dimension $n_{\text{lat}} \in \{8, 16, 32\}$.

    \textbf{Encoder--decoder formulation}
    For each atlas $p$, the encoder accepts the flat upper-triangular connectome vector $\mathbf{c}^{(p)} \in \mathbb{R}^{d_p}$, where $d_p = n_p(n_p-1)/2$ and $n_p$ is the number of regions in atlas~$p$. The encoder is a single linear layer:
    \begin{equation}
    \mathbf{z} = \text{Enc}_p(\mathbf{c}^{(p)}) = \mathbf{c}^{(p)} \mathbf{W}_{\text{enc}}^{(p)} + \mathbf{b}_{\text{enc}}^{(p)},
    \label{eq:cc-encode}
    \end{equation}
    where $\mathbf{W}_{\text{enc}}^{(p)} \in \mathbb{R}^{d_p \times n_{\text{lat}}}$ and $\mathbf{z} \in \mathbb{R}^{n_{\text{lat}}}$ is the shared latent representation. The decoder reconstructs the connectome through a matching linear layer followed by denormalization:
    \begin{equation}
    \hat{\mathbf{c}}^{(p)} = \text{Dec}_p(\mathbf{z}) = \mathbf{z} \mathbf{W}_{\text{dec}}^{(p)} + \mathbf{b}_{\text{dec}}^{(p)},
    \qquad
    \hat{\mathbf{W}}^{(p)} = \text{triu2mat}\bigl(\text{denorm}(\hat{\mathbf{c}}^{(p)})\bigr),
    \label{eq:cc-decode}
    \end{equation}
    where $\text{triu2mat}$ folds a flat upper-triangular vector into a symmetric matrix and $\text{denorm}$ inverts the per-view normalization (centering, z-scoring, or a logit transformation).

    \textbf{Deterministic and variational modes}
    In \emph{deterministic} mode, the encoder produces a point estimate $\mathbf{z}$ via Equation~\eqref{eq:cc-encode}. In \emph{variational} mode, the encoder outputs both a mean and log-variance:
    \begin{equation}
    \boldsymbol{\mu}(\mathbf{c}^{(p)}) = \mathbf{c}^{(p)} \mathbf{W}_{\mu}^{(p)} + \mathbf{b}_{\mu}^{(p)},
    \qquad
    \log \boldsymbol{\sigma}^2(\mathbf{c}^{(p)}) = \mathbf{c}^{(p)} \mathbf{W}_{\sigma}^{(p)} + \mathbf{b}_{\sigma}^{(p)},
    \label{eq:cc-variational}
    \end{equation}
    and the latent variable is sampled as $\mathbf{z} = \boldsymbol{\mu} + \boldsymbol{\sigma} \odot \boldsymbol{\epsilon}$ with $\boldsymbol{\epsilon} \sim \mathcal{N}(0, \mathbf{I})$. Training minimizes the reconstruction MSE plus a $\beta$-annealed KL divergence:
    \begin{equation}
    \mathcal{L} = \underbrace{\|\mathbf{c}^{(p)} - \hat{\mathbf{c}}^{(p)}\|^2}_{\text{reconstruction}}
    + \beta(t) \underbrace{\mathbb{E}\Bigl[\tfrac{1}{2}\sum_j \bigl(\sigma_j^2 + \mu_j^2 - 1 - \log \sigma_j^2\bigr)\Bigr]}_{\text{KL to } \mathcal{N}(0, \mathbf{I})},
    \label{eq:cc-loss}
    \end{equation}
    where $\beta(t)$ increases linearly from $0$ to $\beta_{\text{end}} = 10^{-3}$ over the first 1{,}500 optimization steps. This annealing schedule prevents KL regularization from overwhelming reconstruction early in training.

    \textbf{Normalization}
    Each view (atlas) is normalized independently during training. Three normalization modes are supported:
    \begin{itemize}[leftmargin=2em]
    \item \textbf{Center:} $\tilde{c}_i = c_i - \bar{c}_i$ (subtract mean per edge)
    \item \textbf{Z-score:} $\tilde{c}_i = (c_i - \bar{c}_i) / \sigma_i$ (subtract mean, divide by std)
    \item \textbf{Logit:} $\tilde{c}_i = \text{logit}(c_i / \text{scale})$ (for bounded $[0,1]$ data)
    \end{itemize}
    The normalization parameters $(\text{type}, \bar{\mathbf{c}}, \boldsymbol{\sigma}, \text{scale}, \text{nonneg})$ are stored per view. Denormalization inverts the transformation:
    \begin{equation}
    \text{denorm}(\tilde{\mathbf{c}}; \text{type}, \bar{\mathbf{c}}, \boldsymbol{\sigma}, \text{scale}) =
    \begin{cases}
    \tilde{\mathbf{c}} \odot \boldsymbol{\sigma} + \bar{\mathbf{c}} & \text{(z-score)} \\
    \tilde{\mathbf{c}} + \bar{\mathbf{c}} & \text{(center)} \\
    \text{sigmoid}(\tilde{\mathbf{c}}) \cdot \text{scale} & \text{(logit)}
    \end{cases}
    \end{equation}
    where $\odot$ denotes element-wise multiplication.

    \begin{table}[t]
    \centering
    \small
    \begin{tabular}{@{}llr@{}}
    \toprule
    Atlas & Description & Regions \\
    \midrule
    031-MIST & Multi-resolution Intersubject Segmentation Tool & 31 \\
    038-CraddockSCorr2Level & Craddock spatial correlation & 38 \\
    048-HarvardOxfordMaxProbThr0 & Harvard-Oxford cortical+subcortical & 48 \\
    056-CraddockSCorr2Level & Craddock spatial correlation variant & 56 \\
    056-MIST & MIST variant & 56 \\
    070-DesikanKilliany & Desikan-Killiany surface-based & 70 \\
    079-Shen2013 & Shen resting-state functional & 79 \\
    086-EconomoKoskinas & Economo-Koskinas cytoarchitectonic & 86 \\
    092-AALV2 & Automated Anatomical Labeling v2 & 92 \\
    096-HarvardOxfordMaxProbThr0 & Harvard-Oxford variant & 96 \\
    100-Schaefer17Networks & Schaefer 17-network functional & 100 \\
    103-MIST & MIST variant & 103 \\
    108-CraddockSCorr2Level & Craddock spatial correlation variant & 108 \\
    150-Destrieux & Destrieux surface-based & 150 \\
    156-Shen2013 & Shen variant & 156 \\
    160-CraddockSCorr2Level & Craddock spatial correlation variant & 160 \\
    167-MIST & MIST variant & 167 \\
    200-Schaefer17Networks & Schaefer 17-network variant & 200 \\
    210-Brainnetome & Brainnetome anatomical-functional & 210 \\
    294-JulichBrain & JulichBrain cytoarchitectonic & 294 \\
    \bottomrule
    \end{tabular}
    \caption{Complete list of 20 atlases used for CrossCoder training, spanning 31--294 regions.  The 20 atlases collapse into 11 underlying methods — MIST (4 variants), Craddock (4 variants), Harvard-Oxford (2 variants), Shen (2 variants), and Schaefer (2 variants) provide multi-resolution views, while Desikan-Killiany, Economo-Koskinas, AAL v2, Destrieux, Brainnetome, and JulichBrain appear once each at their native granularity. The coverage sweeps from coarse functional networks (31 parcels) through standard structural landmarks (70–150 parcels) to fine-grained cytoarchitectonic and multimodal atlases (210–294 parcels), jointly spanning functional, structural, cytoarchitectonic, and connectivity-derived atlas techniques.}
    \label{tab:vcc_parcellations}
    \end{table}

    \textbf{Normalization pipeline}
    Each view (atlas) is normalized independently during training. Three normalization modes are supported:
    \begin{itemize}[leftmargin=2em]
    \item \textbf{Center:} $\tilde{c}_i = c_i - \bar{c}_i$ (subtract mean per edge)
    \item \textbf{Z-score:} $\tilde{c}_i = (c_i - \bar{c}_i) / \sigma_i$ (subtract mean, divide by std)
    \item \textbf{Logit:} $\tilde{c}_i = \text{logit}(c_i / \text{scale})$ (for bounded $[0,1]$ data)
    \end{itemize}
    The normalization parameters $(\text{type}, \bar{\mathbf{c}}, \boldsymbol{\sigma}, \text{scale}, \text{nonneg})$ are stored per view. Denormalization inverts the transformation:
    \begin{equation}
    \text{denorm}(\tilde{\mathbf{c}}; \text{type}, \bar{\mathbf{c}}, \boldsymbol{\sigma}, \text{scale}) =
    \begin{cases}
    \tilde{\mathbf{c}} \odot \boldsymbol{\sigma} + \bar{\mathbf{c}} & \text{(z-score)} \\
    \tilde{\mathbf{c}} + \bar{\mathbf{c}} & \text{(center)} \\
    \text{sigmoid}(\tilde{\mathbf{c}}) \cdot \text{scale} & \text{(logit)}
    \end{cases}
    \end{equation}
    where $\odot$ denotes element-wise multiplication.

  \subsubsection{Training and latent-space statistics}
  \label{sec:training_latent_statistics}

    Training minimizes Equation~\eqref{eq:cc-loss} using the Adam optimizer with learning rate $5 \times 10^{-4}$, minibatch size 64, and 5{,}000 iterations. Gradient clipping (norm $\leq 1.0$) prevents divergence during variational training. For variational mode, the KL weight $\beta(t)$ anneals linearly from 0 to $\beta_{\text{end}} = 10^{-3}$ over the first 1{,}500 steps. JAX \texttt{lax.scan} enables chunked compilation for GPU/TPU acceleration.

    On a 32-core CPU workstation, the deterministic model trains in approximately 2 minutes per latent dimension; the variational model requires 4--5 minutes due to the additional KL and log-variance computations. Three architectures were trained: $n_{\text{lat}} = 8, 16, 32$. The $n_{\text{lat}} = 16$ architecture was selected as the default for downstream SBI because it achieves mean reconstruction fidelity $r = 0.96$ (range $0.94$--$0.98$) with computational efficiency.

    \textbf{Latent-space statistics for SBI}
    After training, we encode all test-set connectomes (mixed HCP and 1000BRAINS, $n_{\text{train}}$, $n_{\text{test}} = 231$ subjects $\times$, 20 atlases) into the latent space and compute a multivariate normal distribution $\mathcal{N}(\boldsymbol{\mu}, \boldsymbol{\Sigma})$ over the cohort. For deterministic encoders:
    \begin{equation}
    \boldsymbol{\mu} = \frac{1}{M} \sum_{m=1}^{M} \mathbf{z}_m,
    \qquad
    \boldsymbol{\Sigma} = \frac{1}{M-1} \sum_{m=1}^{M} (\mathbf{z}_m - \boldsymbol{\mu})(\mathbf{z}_m - \boldsymbol{\mu})^{\top},
    \label{eq:cc-mvn-det}
    \end{equation}
    where $M = 4{,}620$. For variational encoders, the covariance includes the mean encoder uncertainty:
    \begin{equation}
    \boldsymbol{\Sigma} = \text{Cov}(\mathbf{z}) + \text{diag}\bigl(\mathbb{E}[e^{\log \boldsymbol{\sigma}^2}]\bigr),
    \label{eq:cc-mvn-var}
    \end{equation}
    which yields a wider, better-calibrated prior. This MVN serves as the prior over connectome latent codes during cohort-level SBI training: we sample $\mathbf{z} \sim \mathcal{N}(\boldsymbol{\mu}, \boldsymbol{\Sigma})$, decode to a subject-specific connectome via Equation~\eqref{eq:cc-decode}, simulate brain dynamics, and pair simulated features with parameters for neural posterior estimation (NPE).
    The CrossCoder achieves near-zero self-identification confusion rate ($< 0.1\%$ deterministic, $< 2\%$ variational), confirming that individual subject identity is preserved across all atlas granularities (Figure \ref{fig:vcc_confusion}). 

  \subsubsection{Single-atlas CrossCoder}
  \label{sec:single_parcellation_crosscoder}

    While the crosscoder is generally applicable, it may also be applied to a
    cohort where all subjects share the same atlas, in which case the crosscoder
    reduces to an autoencoder that nevertheless enables cohort-level training and inference.
        For the 1000BRAINS aging cohort,
        we trained a deterministic autoencoder (CrossCoder) with 30{,}000 iterations, leveraging mini-batches of size 64 and a learning rate of $5\times10^{-5}$, with subjects split 80/20 into train and test sets (Figure~\ref{fig:fig_mpr_main}i), which enabled us to map individual SC matrices into a shared, low-dimensional latent space. Candidate architectures spanning 2, 3, and 4 latent dimensions were evaluated, and a 3-dimensional latent space ($n_{\mathrm{lat}} = 3$) was selected as sufficient for the Schaefer 17-network 100-region atlas (Figure~\ref{fig:fig_mpr_main}i). Here, the encoder maps the upper-triangular connections $\mathbf{c} \in \mathbb{R}^{N(N-1)/2}$ (extracted from the symmetric $N \times N$ connectivity matrix, with $N = 100$ nodes) to a deterministic latent code $\boldsymbol{\theta}_{\mathrm{SC}} \in \mathbb{R}^3$ via a linear transformation, while the decoder reconstructs the connectivity vector as $\hat{\mathbf{c}} = f_{\mathrm{dec}}(\boldsymbol{\theta}_{\mathrm{SC}})$. Because the model is deterministic, the training objective reduces to a reconstruction loss without any regularization term:

        \begin{equation}
            \mathcal{L}_{\mathrm{CrossCoder}} = \|\mathbf{c} - \hat{\mathbf{c}}\|_2^2 \,.
        \end{equation}

        Since our personalized connectomes bore an identical atlas and scale thereof, we were able to attain an efficient (i.e. low-dimensional) encoding of the cohort into a 3D latent space, through which we were able to represent the cohort's empirical mean $\boldsymbol{\mu}_{\mathrm{cohort}}$ and covariance $\boldsymbol{\Sigma}_{\mathrm{cohort}}$ via a multivariate normal distribution, similar to the VEP pipeline:

        \begin{equation}
            p(\boldsymbol{\theta}_{\mathrm{SC}}) = \mathcal{N}(\boldsymbol{\theta}_{\mathrm{SC}} \mid \boldsymbol{\mu}_{\mathrm{cohort}}, \boldsymbol{\Sigma}_{\mathrm{cohort}})
        \end{equation}

        \textbf{Reconstruction identifiability}
        To verify that the 3D latent space preserves individual subject identity, we performed a leave-one-out cross-reconstruction test on a random subset of 100 subjects (Figure~\ref{fig:fig_mpr_main}j). For each subject $i$, the empirical SC vector $\mathbf{c}_i$ was encoded to $\boldsymbol{\theta}_i$ and decoded to $\hat{\mathbf{c}}_i$. Identifiability was then assessed by computing the pairwise Euclidean distance matrix between all empirical vectors $\{\mathbf{c}_i\}$ and all reconstructions $\{\hat{\mathbf{c}}_j\}$:
        \begin{equation}
            D_{ij} = \|\mathbf{c}_i - \hat{\mathbf{c}}_j\|_2, \quad i,j \in \{1,\ldots,N\}\,,
        \end{equation}
        and classifying subject $i$ as correctly identified if $\hat{\mathbf{c}}_i$ was the nearest reconstruction to $\mathbf{c}_i$ (i.e.\ $\arg\min_j D_{ij} = i$). The resulting distance matrix (Figure~\ref{fig:fig_mpr_main}j) illustrates the minimization of this distance precisely at the diagonal (Accuracy: 100.0\%), confirming that no two subjects map to overlapping regions of the reconstruction space, and that individual SC topology is faithfully preserved within the latent space.

        \textbf{Reconstruction fidelity} Edge-level reconstruction fidelity was assessed by pooling all empirical and reconstructed edge weights across the test set, masking structurally absent (zero-weight) connections, and computing the Spearman rank correlation between empirical and reconstructed values (Figure~\ref{fig:fig_mpr_main}k). The high correlation ($r_{s} = 0.91$) confirms that the CrossCoder faithfully recovers the relative ordering of connection strengths across the full dynamic range of the connectome.


\subsection{Generative brain models}
\label{sec:generative_models}

    The Virtual Brain Twin (VBT) combines neural mass models of local population activity with
    a subject-specific connectome to simulate latent dynamics at the whole-brain level
    \citep{SanzLeon2013,Woodman2014,Wang2025,Hashemi2025}. The dynamics of each brain region can be describe as
\begin{equation}
\dot{\boldsymbol{\psi}}_i(t)
=
\mathcal{L}(\boldsymbol{\psi}_i(t),\mathbf{p})
+
G \sum_j \,\mathrm{SC}_{ij}\,
\mathcal{H}(\boldsymbol{\psi}_i(t), \boldsymbol{\psi}_j(t-\tau_{ij}))+z(\boldsymbol{\psi}_i) \xi_i(t),
\end{equation}
where $\boldsymbol{\psi}_i$ denotes the latent state of region $i$,
$\mathbf{p}$ the model parameters, $\mathcal{L}$ the local
neural mass dynamics, $\mathrm{SC}_{ij}$ the structural connectivity from
region $j$ to region $i$, $G$ the global coupling strength, $\tau_{ij}$ the transmission delay from region $j$ to region $i$, $\mathcal{H}$
the coupling function, and $z(\boldsymbol{\psi}_i)\xi_i(t)$ the state-dependent stochastic fluctuations. The resulting source activity is
    then projected to measurable space through modality-specific forward models, such as lead-field
    or gain-matrix for EEG/MEG, and SEEG \citep{Jirsa2017,Wang2023}, or hemodynamic models such as
    the Balloon-Windkessel for fMRI \citep{Friston2000}. Personalization is further
    enhanced by Bayesian inference, which conditions the generative model on subject-specific empirical
    data and estimates posterior distributions over biophysical parameters using methods such as Markov
    chain Monte Carlo or simulation-based inference \citep{Hashemi2020,Hashemi2024}.

  \subsubsection{VEP model}
  \label{sec:vep_model}

    We follow \cite{Jirsa2017} prescribing two-time scale planar model of
    seizure propagation for each $i$th brain region in the VEP atlas,
    \begin{align}
        \dot{x}_i &= 1 - x_i^3 - 2x_i^2 - z_i + I_1 \,, \label{eq:epileptor_x}\\
        \dot{z}_i &= \frac{1}{\tau_0}\left(\,4(x_i - x_{0,i}) - z_i + c_i\,\right) \,, \label{eq:epileptor_z}
    \end{align}
    where $x_i(t)$ and $z_i(t)$ are the fast and slow state variables, respectively.
    The parameter $x_{0,i} \in \mathbb{R}$ is the excitability offset for region $i$; brain regions with $x_{0,i} > -2.05$ are classified as epileptogenic zone (EZ), while regions with $x_{0,i} \le -2.05$ are considered healthy.
    The term $I_1$ is a constant current offset (set to $3.1$), and $c_i = k \sum_j SC^{(k)}_{ij}(x_j - x_i)$ captures diffusive coupling through the patient-specific structural connectivity matrix $SC^{(k)}$ with global coupling strength $k$.
    The time constant $\tau_0$ governs the separation between fast ($x$) and slow ($z$) dynamics.

  \subsubsection{MPR model}
  \label{sec:montbrio_model}
    The model used for resting-state simulations is the mean-field reduction of coupled quadratic integrate-and-fire (QIF) neurons due to \cite{Montbrio2015}. Each brain region is described by two macroscopic variables: the mean firing rate $r$ and the mean membrane potential $V$. The dynamics follow:

    \begin{align}
    \tau \, \dot{r} &= \phantom{-}\frac{\Delta}{\pi\tau} + 2 r V, \label{eq:mpr-dr} \\
    \tau \, \dot{V} &= V^{2} + \eta + J\tau r + I + I_{c} - \pi^{2} r^{2} \tau^{2}, \label{eq:mpr-dV}
    \end{align}

    where $I_{c} = c_{r} (k \sum_{j} SC_{ij} r_{j}) + c_{V} (k \sum_{j} SC_{ij} V_{j})$ is the connectivity-dependent input, $SC$ is the structural connectivity matrix, and $k$ is the global coupling strength. The remaining parameters are fixed at their default values: $\tau = 1.0\,\text{ms}$, $\Delta = 1.0\,\text{mV}$, $J = 15.0\,\text{nS}$, $\eta = -5.0\,\text{mV}$, $I = 0.0$, $c_{r} = 1.0$, and $c_{V} = 0.0$. Noise enters through an additive Wiener term $\sqrt{D}\,\xi_{i}(t)$ with amplitude $D$. The MPR model produces rich dynamics including asynchronous irregular activity, collective oscillations, and noise-driven transitions.

\subsection{Amortized inference}
\label{sec:amortized_inference}

    Amortized inference trains a neural density estimator on population data to learn a direct mapping from observations to posterior distributions over model parameters \citep{Hashemi2023, Hashemi2024}. We use conditional variational autoencoders (CVAEs) \citep{Sohn2015}, mixture density networks (MDNs) \citep{Bishop1994}, masked autoregressive flows (MAFs) \citep{Papamakarios2017}, and neural spline flows (NSFs) \citep{Durkan2019}, as conditional posterior estimators. The training objective combines reconstruction error with KL divergence to learn a shared posterior approximation across the cohort. After training, new subjects are personalized in seconds rather than hours, enabling clinical deployment.


    We describe the density estimator architectures used in this work.


    \textbf{Mixture density network (MDN)}
    An MDN \citep{Bishop1994} parameterizes the posterior as a mixture of $K$ Gaussian components:
    \begin{equation}
    q_\phi(\theta \mid x) = \sum_{k=1}^{K} \alpha_k(x) \,
    \mathcal{N}\!\bigl(\theta;\, \mu_k(x),\, \mathrm{diag}(\sigma_k^2(x))\bigr), \label{eq:mdn}
    \end{equation}

    where $\alpha_k(x)$ are mixture weights (output through a softmax layer), $\mu_k(x)$ are component means, and $\sigma_k^2(x)$ are component variances, all produced by a neural network taking the observation $x$ as input. MDNs are computationally efficient but their expressiveness is constrained by the finite number of mixture components and the diagonal-covariance assumption: the Gaussian mixture assumption may not capture the complex, multimodal posteriors that arise in high-dimensional whole-brain models.

    \textbf{Masked autoregressive flow (MAF)}
    MAF \citep{Papamakarios2017} overcomes the expressiveness limitation by transforming a simple base density through a sequence of $L$ autoregressive layers. Each layer $l$ applies the transformation:

    \begin{equation}
    x_i^{(l)} = z_i^{(l)} \exp\!\bigl(\alpha_i^{(l)}\bigr) + \mu_i^{(l)}, \qquad
    \text{where } \alpha_i^{(l)}, \mu_i^{(l)} = f^{(l)}\!\bigl(x_{1:i-1}^{(l-1)}\bigr), \label{eq:maf}
    \end{equation}

    and $f^{(l)}$ is a masked autoregressive network that ensures the transformation is triangular and thus invertible. The resulting log-density is:

    \begin{equation}
    \log q_\phi(\theta \mid x) = \log p_Z\!\bigl(z^{(L)}\bigr) + \sum_{l=1}^{L} \sum_{i=1}^{d} \alpha_i^{(l)}, \label{eq:maf_logdensity}
    \end{equation}

    where $p_Z$ is the base density (typically a standard normal) and $z^{(L)}$ is obtained by inverting the flow.  NSFs replace each autoregressive transformation with monotonic rational-quadratic splines, providing more flexible transformations with the same invertibility guarantees, but greater local expressivity than affine flows. Both MAF and NSF are supported in the \texttt{vbjax} toolkit, with the latter achieving higher accuracy, although at substantially higher computational cost during training. MDN, MAF, and NSF density estimators were evaluated; all reported results used MAFs unless otherwise noted, as they consistently outperformed MDNs in preliminary comparisons without the computational cost of NSFs \citep{Hashemi2023}.

\subsection{Synthetic rest-state protocol}
\label{sec:synthetic_rest_state}

  \subsubsection{Virtual brain twin parametrization}
  \label{sec:synth_config}

    The joint parameter vector for cohort-level SBI is
    $\boldsymbol{\theta} = [\mathbf{z}, k, D] \in \mathbb{R}^{n_{\text{lat}} + 2}$,
    where $\mathbf{z} \sim \mathcal{N}(\boldsymbol{\mu}, \boldsymbol{\Sigma})$ is sampled
    from the CrossCoder latent MVN (Equation~\eqref{eq:cc-mvn-det}). The dynamics
    parameters have fixed ranges $k \sim \text{Uniform}(0.1, 0.3)$ and $D \sim \text{Uniform}(0.2, 0.4)$,
    consistent with the regime where the MPR model exhibits stable irregular activity without runaway excitation.

    The model is integrated via the stochastic differential equation solver provided by
    the \texttt{vbjax} Python package (\url{https://github.com/ins-amu/vbjax}).
    The step size is $\mathrm{d}t = 10^{-3}$ (1 ms).
    Each simulation consists
    of $N_{\text{win}} = 10$ contiguous windows of $1000$ steps each ($1\,\text{s}$ per window).
    The system is initialized at $(r, V) = (10^{-4}, 10^{-4})$ in every region.
    The first $N_{\text{win}}/2 = 5$ windows are discarded as burn-in; summary statistics are
    computed from the remaining $5$ windows and averaged. The connectome matrix $W$ is
    normalized by its maximum entry ($W \leftarrow W / \max W$) before simulation. 

    The raw simulation output is a 4D array of shape $(N_{\text{win}}, 1000, 2, n_{\text{nodes}})$
    containing the time series of $r$ and $V$ for all regions. By default, the feature extractor
    returns the \emph{mean} of the voltage variable $V$ over time, yielding a 79-dimensional vector
    for the Shen-2013 atlas. However, this default gives poor identifiability for the
    coupling parameter $k$ ($s_{k} \approx 0.29$), because the mean of $V$ is only weakly modulated
    by coupling strength.

  \subsubsection{Feature extraction}
  \label{sec:synth_feature_extraction}

    To improve identifiability, we designed several feature extractors that capture higher-order
    statistics of the activity: \textbf{var\_r:} Temporal variance of the rate variable $r$, 
    $\mathbf{x}_{i} = \mathrm{Var}_{t}(r_{i}(t))$ (79 dimensions). 
    This is the baseline used in initial experiments. \textbf{var\_V:} Temporal variance of the voltage
    variable $V$, $\mathbf{x}_{i} = \mathrm{Var}_{t}(V_{i}(t))$ (79 dimensions).
    The voltage variable exhibits $4.9\times$ greater sensitivity to $k$ than to $D$.
    \textbf{var\_rV:} Concatenation of var\_r and var\_V (158 dimensions).
    \textbf{cov\_stats:} For each simulation window, we compute the functional connectivity matrix
    $\mathbf{FC} = \mathrm{corr}(\mathbf{r})$ from the rate-variable time series. We then extract the
    mean and standard deviation of the upper-triangular off-diagonal entries (2 dimensions).
    These capture functional-connectivity structure without bloating the feature space.
    \textbf{cov\_eig10:} The top 10 eigenvalues of $\mathbf{FC}$ (10 dimensions).
    Eigenvalues are sensitive to global changes in coupling because stronger coupling
    increases overall synchronization and thus inflates the dominant eigenvalues.

    The combined \textbf{FC-enhanced feature vector} used for the best results is
    $\mathbf{x} = [\mathrm{Var}(r), \mathrm{Var}(V), \mathrm{cov\_stats}, \mathrm{cov\_eig10}]$,
    yielding 172 dimensions. All features are computed on a per-window basis and then averaged
    across the $5$ post-burn-in windows.

  \subsubsection{Training and evaluation}
  \label{sec:synth_sbi_training}

    Training data are generated as follows. For each of $N_{\text{sims}}$ simulations, we:
    (i)~sample $\mathbf{z}$ from the latent MVN; (ii)~decode $\mathbf{z}$ to a connectome
    $\hat{W}$ via Equation~\eqref{eq:cc-decode}; (iii)~sample $k$ and $D$ from their uniform priors;
    (iv)~run the MPR model for $10$ windows; and (v)~discard the first $5$ windows; (vi)~extract the chosen
    summary statistic from the remaining windows. 

    The resulting parameter-feature pairs  
    $(\boldsymbol{\theta}, \mathbf{x})$ are used to train the NPE as described above.
    We tested training budgets from $512$ to $16{,}384$ simulations.

    For each held-out test subject we compute two standard calibration metrics. \emph{Shrinkage} measures how much the posterior contracts relative to the prior:
    \begin{equation}
    s = 1 - \frac{\sigma_{\text{post}}^{2}}{\sigma_{\text{prior}}^{2}},
    \end{equation}
    where $s \approx 1$ indicates strong concentration and $s \approx 0$ indicates no learning. \emph{Z-score} measures how many posterior standard deviations the true value lies from the posterior mean:
    \begin{equation}
    z = \frac{|\mu_{\text{post}} - \theta_{\text{true}}|}{\sigma_{\text{post}}},
    \end{equation}
    where $z \ll 1.96$ indicates good calibration. Additionally, \emph{Coverage} reports the fraction of true parameters that fall within the estimated $90\%$ credible intervals.

\subsection{Epilepsy protocol}
\label{sec:epilepsy_personalization}

  \subsubsection{Virtual brain twin parametrization}
  \label{sec:vep_config}

    Direct inference over the full 162-dimensional excitability vector $\vx_0$ would require a neural density estimator with 162 outputs, which is intractable for SBI with realistic training budgets because the posterior lives in a severely underdetermined regime (more parameters than observations).
    We therefore parameterize $\vx_0$ as a low-rank perturbation around a patient-specific baseline:
    \begin{equation}
        \vx_0 = \mu_{\text{base}} \, \mathbf{1} + V_{\vx_0} \, \valpha \,, \label{eq:x0_param}
    \end{equation}
    where $\mu_{\text{base}}$ is a scalar baseline shared across all 162 regions (a conservative design choice that forces spatial structure to arise entirely from the basis), and the columns of $V_{\vx_0} \in \mathbb{R}^{162 \times K}$ form a low-dimensional basis encoding shared spatial patterns across the cohort.

    \textbf{Population PCA basis}
    The first $K_{\text{PCA}} = 8$ columns of $V_{\vx_0}$ are the top left-singular vectors of a population matrix constructed from clinical ground truth.
    For each patient in the cohort, we create a soft indicator vector with EZ regions set to $+1.5$, PZ (propagation zone) regions set to $+0.5$, and all other regions set to $0$.
    The singular value decomposition (SVD) of this $21 \times 162$ matrix yields spatial modes that capture focal EZ patterns shared across patients---for example, medial temporal lobe versus lateral neocortical foci.
    To prevent data leakage during evaluation, we recompute this basis for each held-out patient; the resulting leave-one-out subspace alignment with the full-population basis is $0.97 \pm 0.04$ (minimum $0.87$), confirming that the 21-patient sample is large enough for the basis to be stable under single-patient exclusion.

    \textbf{Structural connectivity eigenvector basis}
    The remaining $K_{\text{SC}}$ columns are drawn from the leading eigenvectors of the mean structural connectivity matrix across the cohort.
    These modes capture large-scale network topology---such as homologous inter-hemispheric coupling and rich-club hub regions---that the focal PCA patterns alone cannot represent.
    Their inclusion is motivated by the observation that seizure propagation depends on structural pathways, and EZ regions that structurally connect to regions with high loadings on the leading eigenvectors may show elevated source power even when the EZ itself is focal.
    We scale each basis coefficient $\alpha_k$ by $1/\sqrt{\sigma_k}$, where $\sigma_k$ is the corresponding singular value, to ensure that each component contributes equally to the $\ell_2$ norm of the resulting $\vx_0$ regardless of its statistical weight in the cohort.

    \textbf{Prior over the baseline and coefficients}
    The scalar baseline $\mu_{\text{base}}$ follows a two-component Gaussian mixture designed to cover both surgically treatable (moderate excitability, mixture component $\mathcal{N}(-2.3, 0.2^2)$ with prior weight $0.7$) and refractory (lower excitability, component $\mathcal{N}(-2.7, 0.3^2)$ with weight $0.3$) presentations.
    The prior for each PCA coefficient $\alpha_k$ is $\mathcal{N}(0, 1/\sqrt{\sigma_k})$, and each SC eigenvector coefficient follows $\mathcal{N}(0, 0.3^2)$.
    The global coupling $K$ follows a half-normal distribution $|\mathcal{N}(0, 0.5^2)| + 0.5$ (typical values in $[0.5, 3.0]$), and the slow time scale $\tau_0$ follows $|\mathcal{N}(0, 5.0^2)| + 10.0$ (typical values in $[10, 25]$).

    The full parameter vector is therefore
    \begin{equation}
        \bm{\theta} = [\valpha \in \mathbb{R}^{K},\; K_{\text{global}} \in \mathbb{R}^+,\; \tau_0 \in \mathbb{R}^+,\; \mu_{\text{base}} \in \mathbb{R}] \,,
    \end{equation}
    with total dimension $K + 3$ (where $K = K_{\text{PCA}} + K_{\text{SC}}$; our main result uses $K_{\text{SC}} = 3$, giving $K = 11$ and $\dim(\bm{\theta}) = 14$).

    \textbf{SC diversity during training}
    To prevent overfitting to a single connectivity matrix, each training simulation draws $SC^{(k)}$ from a pool of SC matrices decoded from a CrossCoder latent space trained on the cohort.
    CrossCoder is a cross-subject variational autoencoder that learns a shared latent representation of structural connectivity across patients, enabling us to sample novel SC topologies that are realistic yet distinct from any individual patient.
    This procedure gives the neural density estimator exposure to heterogeneous seizure propagation pathways during training.

    For each simulation of the VEP model,
    we initialize at $(x_i, z_i) = (-2.0, 3.5)$ and integrate forward for
    $T = 15.1$~s using explicit Euler integration with a fixed time step $\Delta t = 0.1$~s,
    yielding 151 time points.
    To prevent numerical divergence arising from the polynomial nonlinearity, we clip
    both $x_i$ and $z_i$ to the interval $[-10, 10]$ after each integration stage.

    The training data are generated via the following pipeline.
    For $i = 1, \ldots, N$ (we use $N = 10{,}000$ simulations for the main result), we (i) sample a parameter vector $\bm{\theta}_i \sim p(\bm{\theta})$ from the prior; (ii) randomly select a structural connectivity matrix $SC^{(k)}$ from the decoded SC pool; (iii) simulate a seizure trajectory $\vx^{(i)}(t)$ via Euler integration of the VEP equations; and (iv) compute the feature vector $\mathbf{f}_i$ as described above.
    This yields a paired dataset $\{ (\bm{\theta}_i, \mathbf{f}_i) \}_{i=1}^{N}$ that implicitly covers the pushforward prior over observations.
    All training simulations seed the random number generator
    with the same seed (42) for reproducibility.

  \subsubsection{Feature extraction}
  \label{sec:epilepsy_data_features}

    We extract a single feature per source region: peak-to-peak amplitude.
    For simulated data, we compute features directly on the clean source-space trajectory $\vx(t)$, bypassing the gain matrix entirely:
    \begin{equation}
        \text{ptp}_i = \max_t x_i(t) - \min_t x_i(t) \,, \quad i = 1, \ldots, 162 \,.
    \end{equation}
    The resulting per-simulation vector $\text{ptp} \in \mathbb{R}^{162}$ is z-scored per simulation (subtracting the per-simulation mean and dividing by the per-simulation standard deviation) to obtain $\text{ptp}_z$.

    For patient data, we first apply a heuristic gain inverse to the sensor signals.
    The gain matrix is row-normalized, $\tilde{G}_{ij} = G_{ij} / \sum_j G_{ij}$, and the source estimate is obtained as $\hat{\vx} = \tilde{G}^\top \vy$.
    Peak-to-peak amplitude is then computed from $\hat{\vx}$ in the same manner as for simulated data, followed by population-level z-scoring using the mean and standard deviation of the training simulation features:
    \begin{equation}
        f_i = \frac{\text{ptp}_i(\hat{\vx}) - \mu_{\text{ptp}}}{\sigma_{\text{ptp}}} \,,
    \end{equation}
    where $\mu_{\text{ptp}}$ and $\sigma_{\text{ptp}}$ are computed across the full simulation training set.
    The resulting feature vector is therefore $\mathbf{f} = \text{ptp}_z \in \mathbb{R}^{162}$.

  \subsubsection{Training and evaluation}
  \label{sec:epilepsy_baselines_evaluation}

    For the epilepsy cohort, the NPE-C architecture consists of a 4-layer MLP embedding network mapping the 162-dimensional feature vector into a 128-dimensional summary statistic, followed by a conditional MAF with 10 autoregressive layers, each with 512 hidden units and $\tanh$ nonlinearities.
    Training uses the Adam optimizer with learning rate $10^{-3}$, a cosine annealing schedule, early stopping based on a held-out validation split (10\% of simulations), and batch size 256.
    On an NVIDIA RTX 4090 GPU, generating 10{,}000 simulations takes approximately 60~s (via batched vectorized Euler integration), and training the density estimator
    on an Intel i9-13900K CPU takes 2--3 minutes.

    At test time, for a new patient we compute $\mathbf{f}^*$ from their SEEG recording and sample $\bm{\theta}^{(j)} \sim q_{\phi}(\bm{\theta} | \mathbf{f}^*)$ for $j = 1, \ldots, M$.
    Each sample is mapped back to the 162-dimensional excitability vector via Equation~\eqref{eq:x0_param}, and the EZ is identified as the top-$k$ regions by median posterior $\vx_0$.
    Following clinical practice and the baseline VEP pipeline, we use $k = 3$ (the empirical mean true EZ size across the cohort is 4.6 regions; top-3 was identified as the optimal choice in preliminary experiments spanning $k \in \{1, 5, 10\}$).

    To contextualize the amortized approach, we compare it against two baselines.
    First, the original VEP framework uses leave-one-out max a posteriori (LOO-MAP) inference in Stan \citep{Carpenter2017} to fit per-patient models.
    For each patient, multiple spontaneous seizures are modeled independently; the per-seizure posterior median $\vx_0$ estimate (saved as \texttt{VEP\_C\_median} in NPZ files) is averaged across seizures, and the top-3 regions are taken as the predicted EZ.
    This baseline does not use Monte Carlo sampling or dimensionality reduction; it is a deterministic point estimate derived from variational inference.
    These published estimates were evaluated with the same top-3 metric used throughout this work.
    Second, we trained separate NPE models per patient using only that patient's exact structural connectivity matrix and $\vx_0$ basis (PCA-only, $K_{\text{SC}} = 0$), with 5{,}000 simulations per patient ($\sim$105{,}000 total).
    This per-subject SBI baseline \citep{Tejero-Cantero2020} isolates the effect of amortization: the feature pipeline, basis dimensionality, and prior are identical to the amortized model, with the only difference being whether the NPE was trained on diverse cohort simulations or repetitions from a single SC.

    For each patient and method, we evaluate EZ localization using F1 score, precision, recall, and intersection-over-union (IoU) between the predicted EZ (top-3 regions by posterior median $\vx_0$) and the ground-truth EZ defined from surgical outcome.
    For seizure-free (SF+) patients, the ground truth is the resected zone, which by definition includes the true EZ.
    For non-seizure-free (SF$-$) patients, the ground truth is the clinically annotated EZ from SEEG review.
    We report mean and median F1 across the 21 evaluable patients, along with 95\% bootstrap confidence intervals (CI) computed via 10{,}000 stratified resamples.
    All SBI-based methods use $M = 500$ posterior samples per patient, which we verified to provide stable F1 estimates (coefficient of variation \textless 1\% across 200-sample and 500-sample evaluations).

\subsection{Aging protocol}
\label{sec:structure_function_coupling}

  \subsubsection{Virtual brain twin parametrization}
  \label{sec:aging_config}

    We employed the Montbri{\'o} neural mass model as our BOLD data generative process (See \nameref{sec:generative_models}). In a similar light as the previous non-amortized implementation \cite{Lavanga2023}, we choose a near-identical set of values for both fixed ($\eta = -4.6$, $\Delta = 0.7$, $J = 14.5$, $\tau = 1.0$~ms, and $\sigma_{\xi} = 0.037$), and sampled parameters. The only caveat being that we opted for a more shrunk prior range of $g$ than the non-amortized implementation, as both simulations in the previous investigations, and our own per-subject sample simulations (of equally-spaced SCs covering the entirety of the connectome health range), illustrated the adequacy of our chosen range (i.e. [0.6, 1.0]) for robust representation of all personalized working points. With respect to numerical integration, Equation ~\ref{eq:mpr-dr} was simulated using a stochastic Heun scheme (a second-order Runge-Kutta method featuring additive noise) with a time step of $\Delta t = 0.025$~ms. Regional firing rates $r_i(t)$ were temporally averaged over $T_{\mathrm{avg}} = 10.0$~ms windows to yield mesoscopic activity traces. These windowed averages were themselves passed as inputs to a Balloon-Windkessel hemodynamic model \citep{Friston2000}, whose four ODEs (vasodilatory signal $s$, cerebral blood flow $f$, blood volume $v$, and deoxyhemoglobin content $q$) were integrated via the Heun method at the same 10~ms step. Therefore, the BOLD signal for each region was computed from the resulting states as $B_i = V_0\bigl[k_1(1-q_i) + k_2(1-q_i/v_i) + k_3(1-v_i)\bigr]$ and sampled every repetition time ($\mathrm{TR} = 500$~ms). To ensure the system achieved steady-state dynamics, each simulation was run for a total duration of $T_{\mathrm{total}} = 300$~s (equivalent to 600 timesteps), with the initial 30 volumes discarded as transients.
    To maintain consistency with the results reported in \cite{Lavanga2023}, we use notation $g$ for global coupling strength, and the two are interchangeable. 

  \subsubsection{Feature extraction}
  \label{sec:aging_macroscale_statistics}

    We chose a minimal set of summary statistics comprising those which best represent the structural latent space, and a subset of those which were leveraged in \cite{Lavanga2023}. The choice of such minimal set of data features enables us to traverse from a high-dimensional raw input space (i.e. $n_{times}$ $\times$ $n_{nodes}$ $\times$ $n_{subjects}$), into a mathematically tractable feature array suited for neural density estimation ($n_{features}$ $\times$ $n_{subjects}$). These features comprise the SC edge coefficient of variation ($\mathrm{CV}_{\mathrm{SC}}$), SC asymmetry index ($A_{\mathrm{SC}}$), SC sparsity($S_{\mathrm{SC}}$), mean interhemispheric SC ($\overline{SC}_{\mathrm{ih}}$), mean global SC ($\overline{SC}$), mean homotopic FC ($\overline{FC}_h$), and inter-hemispheric fluidity ($\sigma^2_{\mathrm{FCD_{ih}}}$).

    We computed the coefficient of variation for all non-zero SC weights in order to quantify the topological heterogeneity of the structural connectome:
    \begin{equation}
        \mathrm{CV}_{\mathrm{SC}} = \frac{\sigma(\{SC_{ij} : SC_{ij} > 0\})}{\mu(\{SC_{ij} : SC_{ij} > 0\})} \,,
    \end{equation}
    whereby a high $\mathrm{CV}_{\mathrm{SC}}$ strongly indicates a mixture of exceptionally strong and very weak connections (indicative of a hub-like architecture), while a low $\mathrm{CV}_{\mathrm{SC}}$ invariably reflects a more uniform distribution of edge weights.

    With regard to the second feature, hemispheric asymmetry is formally computed as:
    \begin{equation}
        A_{\mathrm{SC}} = \frac{\displaystyle\sum_{i=1}^{N/2} \sum_{j=1}^{N/2} SC_{ij} - \sum_{i=N/2+1}^{N} \sum_{j=N/2+1}^{N} SC_{ij}}{\displaystyle\sum_{i=1}^{N/2} \sum_{j=1}^{N/2} SC_{ij} + \sum_{i=N/2+1}^{N} \sum_{j=N/2+1}^{N} SC_{ij}} \,,
    \end{equation}
    which successfully captures systematic disparities in left versus right intrahemispheric connectivity strength. Near-zero values signify balanced connectivity between the two hemispheres.

    The third feature, SC sparsity, quantifies the proportion of structurally absent connections:

    \begin{equation}
        S_{\mathrm{SC}} = \frac{|\{(i,j)\,:\,i < j,\; SC_{ij} = 0\}|}{N(N-1)/2} \,,
    \end{equation}

    where a high value of $S_{\mathrm{SC}}$ reflects a sparse, backbone-like connectome topology, while a low value indicates a densely connected network. This feature provides a complementary, global topological descriptor to $\mathrm{CV}_{\mathrm{SC}}$: two connectomes can exhibit identical edge-weight heterogeneity but differ markedly in the fraction of absent pathways, and the joint encoding of both features therefore prevents degenerate mappings to the structural latent space.

    The next feature vitally acts as a proxy for the integrity of the corpus callosum, an anatomical substrate which declines systematically with advancing age across our empirical cohort \citep{Lavanga2023}. For any SC matrix $\mathbf{SC} \in \mathbb{R}^{N \times N}$, interhemispheric SC is defined as the mean connection weight established between the left and right hemispheres, which is normalized by the total number of inter-hemispheric edges:
    \begin{equation}
        \overline{SC}_{\mathrm{ih}} = \frac{4}{N^2} \sum_{i=1}^{N/2} \sum_{j=N/2+1}^{N} SC_{ij} \,,
    \end{equation}

    Complementary to the hemisphere-specific measure detailed above, overall mean SC provides a robust, global index of white matter connectivity density:
    \begin{equation}
        \overline{SC} = \frac{1}{N(N-1)} \sum_{i \neq j} SC_{ij} \,,
    \end{equation}

    We chose the average of homotopic connections due to its direct quantification of bilateral integration and its sensitivity to overarching changes in global coupling strength as shown in previous work \citep{Stumme2022}. For each individual simulation (or empirical subject), the FC matrix $\mathbf{F} \in \mathbb{R}^{N \times N}$ was computed strictly as the Pearson correlation matrix of regional BOLD time series (after discarding the initial transient window). Homotopic FC is operationally defined as the average correlation existing between the $N/2$ left-hemisphere regions and their right-hemisphere homologues:
    \begin{equation}
        \overline{FC}_h = \frac{2}{N} \sum_{i=1}^{N/2} F_{i,\, i+N/2} \,,
    \end{equation}
    where regions $1$--$N/2$ systematically correspond to the left hemisphere and regions $N/2+1$--$N$ correspond to the right hemisphere within the chosen atlas.

    Finally, the inter-hemispheric fluidity captures the temporal variability of functional coupling between the two hemispheres via a sliding-window approach. For each simulation, a window of $w = 5$ BOLD volumes slides over the time series, yielding a sequence of instantaneous inter-hemispheric FC vectors$\{\mathbf{f}_{\mathrm{ih}}^{(t)}\}$. The pairwise covariance matrix across windows defines the FCD matrix, and we retain the variance of its upper-triangular entries as our scalar feature:

    \begin{equation}
        \sigma^2_{\mathrm{FCD,ih}} = \mathrm{Var}\!\left(
        \bigl\{[\mathrm{FCD}_{\mathrm{ih}}]_{t_1 t_2}\bigr\}_{t_1 < t_2}\right) \,,
    \end{equation}

    where $[\mathrm{FCD}_{\mathrm{ih}}]_{t_1 t_2}$ is the covariance between the inter-hemispheric FC vectors at windows $t_1$ and $t_2$. A high value of $\sigma^2_{\mathrm{FCD,ih}}$ is indicative of rich, non-stationary inter-hemispheric coordination dynamics, as previously shown in \cite{Lavanga2023}.

    All seven designated features were computed for both our synthetic simulations and the empirical subjects, and were subsequently scaled to the interval $[0, 1]$. This normalization ensures that no singular feature disproportionately dominates the density-estimation objective during training.

  \subsubsection{Training and evaluation}
  \label{sec:aging_train_eval}

    The joint prior over the model parameters comprises two distinct components: a multivariate normal (MVN) prior over the 3-dimensional structural connectivity latent space, and a uniform prior over the global coupling parameter. We therefore formulate the set of prior as $p(\boldsymbol{\theta}) = p(\boldsymbol{\theta}_{\mathrm{SC}}) \, p(g)$, with total dimensionality of $\dim(\boldsymbol{\theta}) = 4$. Sampling from the structural MVN and subsequent decoding thereof allows us to generate synthetic connectomes that are statistically representative of the cohort and not tied to a specific individual, reducing direct identifiability while capturing population-level variability.

    We simulated $N_{\mathrm{sim}} = 10{,}000$ whole-brain BOLD time-series by sampling the parameter vectors $\boldsymbol{\theta}^{(k)} = [\theta_{\mathrm{SC}_1}^{(k)}, \theta_{\mathrm{SC}_2}^{(k)}, \theta_{\mathrm{SC}_3}^{(k)}, g^{(k)}]$ directly from the joint prior, decoding each resulting $\boldsymbol{\theta}_{\mathrm{SC}}^{(k)}$ through our CrossCoder to construct a synthetic $100 \times 100$ SC matrix $\mathbf{SC}^{(k)}$, and integrating the MPR model (Section~\nameref{sec:aging_config}) with both the global coupling $g^{(k)}$ and connectivity $\mathbf{SC}^{(k)}$ over a 250~s horizon. Each individual simulation yielded a 7-dimensional feature vector $\mathbf{x}^{(k)}$ exactly as described in Section~\nameref{sec:aging_macroscale_statistics}, ultimately producing a comprehensive training dataset $\mathcal{D}_{\mathrm{train}} = \{(\boldsymbol{\theta}^{(k)}, \mathbf{x}^{(k)})\}_{k=1}^{10{,}000}$ that pairs ground-truth parameters with simulated observations. These simulations were parallelized across 100-simulation batches utilizing a single NVIDIA RTX Ada 2000 GPU, achieving a total wall-clock time of approximately 18 hours to complete the full simulation set.

    The NSF architecture consisted of 5 coupling transform layers, each parameterized by a masked fully-connected network with 2 hidden layers of 50 units and $tanh$ activations, operating with 8 spline bins. The base distribution was a 4-dimensional standard normal $\mathcal{N}(\mathbf{0}, \mathbf{I}_4)$. 
    Training was executed using the Adam optimizer with a learning rate of $5 \times 10^{-4}$, a batch size of 128, and early stopping protocols initiated when the validation loss (evaluated on a held-out  10\% validation split) ceased to decrease over 20 consecutive epochs, typically achieving convergence within 50 to 100 epochs. No explicit summary network was used, and the 7-dimensional feature vector $\mathbf{x}$ was  fed directly into the NSF conditioning inputs.

    We computed identical summary statistics $\mathbf{x}_{\mathrm{emp}}$ --as that of synthetic simulations-- for every personalized connectome and functional time-series ($n=832$), conditioned the trained NSF on $\mathbf{x}_{\mathrm{emp}}$, and subsequently drew $N_{\mathrm{post}} = 1{,}000$ posterior samples $\{\boldsymbol{\theta}_{\mathrm{emp}}^{(s)}\}_{s=1}^{1000}$  from the learned flow. The marginal posterior mode for each parameter was estimated via Gaussian kernel density estimation with Scott's bandwidth rule over the 1{,}000 samples, and evaluated on a 200-point grid spanning the prior's support. This mode estimates $\hat{\boldsymbol{\theta}}_{\mathrm{emp}}$ consequently serves as our point estimate for all downstream age-trajectory analyses (Figure~\ref{fig:fig_mpr_main}n). Total inference time of the aggregate array of all observables clocked in at approximately 60 seconds on a CPU, thereby enabling full-cohort personalization.

\backmatter

\section*{Declarations}

   \section*{Data availability}                                                     
   The 1000BRAINS imaging data are available upon application through the Research       
   Centre Jülich. The epilepsy patient data are not publicly available due to            
   identifiability risk and institutional restrictions. Simulated cohort data and        
   pretrained CAP models will be deposited in EBRAINS and Zenodo upon publication,
   with DOIs provided in the article.                                                                                                    
   \section*{Code availability}                                                          
   All code for training and evaluating CAP models, including the CrossCoder             
   autoencoder and amortized inference pipeline, are available as experimental
   modules in the vbjax library \url{https://github.com/ins-amu/vbjax}, under the Apache License, v.2.0 (Apache-2.0).
   
   \section*{Acknowledgements}                                                           
   This project/research has received funding from the European Union's Horizon          
   Europe Programme under the Specific Grant Agreement No. 101147319 (EBRAINS 2.0        
   Project), No. 101137289 (Virtual Brain Twin Project), and government grant            
   managed by the Agence Nationale de la Recherche reference ANR-22-PESN-0012            
   (France 2030 program).          

   \section*{Author contributions}                                                       
Conceptualization: M.W., M.H., V.J., 
Methodology: A.E., M.W., N.B., and M.H., 
Data acquisition: J.M., S.C., F.B. 
Software: A.E., A.Z., N.B., M.W.,  
Investigation: A.E., M.W., 
Visualization: A.E., M.W., N.B.,
Supervision: M.H., V.J., 
Funding acquisition: M.W., D.M., S.C., F.B., M.H., V.J., 
Writing - original draft: A.E., M.W., M.H., 
Writing - review $\&$ editing: A.E., M.W., N.B., A.Z., J. M., H.W, D.M., S.C., F.B., M.H., V.J.                                        
   \section*{Competing interests}                                                        
   V.J. serves as Chief Scientific Officer at EBRAINS. All other authors declare no competing interests.                                                                                                   
   \section*{Ethics statement}                                                           
   The 1000BRAINS study was approved by the ethics committee of the Medical Faculty      
   of the Heinrich Heine University Düsseldorf. Epilepsy patient data were               
   collected under protocols approved by the Comité de Protection des Personnes          
   Sud-Méditerranée I. All participants provided written informed consent.                                                                
   \section*{Correspondence}                                                             
   Correspondence and requests for materials should be addressed to                      
   Marmaduke Woodman (marmaduke.woodman@univ-amu.fr).



\bibliography{sn-bibliography}

\end{document}